\documentclass[aps,prl,twocolumn,superscriptaddress,10pt,article,showpacs]{revtex4-2}

\usepackage{blindtext}
\usepackage{lipsum}
\usepackage{graphics}
\usepackage{amsmath}
\usepackage{graphicx}
\usepackage{graphics}
\usepackage{amssymb}
\usepackage{verbatim}
\usepackage{physics}
\usepackage{float}
\usepackage[normalem]{ulem}
\usepackage[dvipsnames]{xcolor}
\usepackage{bbm}
\usepackage{bm}
\usepackage{enumitem} 

\definecolor{blueviolet}{rgb}{0.2, 0.2, 0.6}
\definecolor{webgreen}{rgb}{0,.5,0}
\definecolor{webbrown}{rgb}{.6,0,0}
\usepackage[bookmarks=false,
	colorlinks=true, 
	urlcolor=webbrown, 
	linkcolor=blueviolet, 
	citecolor=webgreen,
	pdfstartpage=1,
	pdfstartview={FitH},  
	bookmarksopen=false
	]{hyperref}

\makeatletter 
    
\renewcommand\onecolumngrid{
\do@columngrid{one}{\@ne}%
\def\set@footnotewidth{\onecolumngrid}
\def\footnoterule{\kern-6pt\hrule width 1.5in\kern6pt}%
}

\newcommand{\be}{\begin{equation}}
\newcommand{\ee}{\end{equation}}
\newcommand{\bs}{\begin{split}}
\newcommand{\es}{\end{split}}

\renewcommand{\d}{\partial}

\newcommand{\calZ}{\mathcal{Z}}
\renewcommand{\o}{\omega}
\newcommand{\R}{r}

\begin{document}

\title{Cooling the Sachdev-Ye-Kitaev model using thermofield double states}

\author{Thomas Schuster}
\thanks{These authors contributed equally to this work.}
\affiliation{Walter Burke Institute for Theoretical Physics, California Institute of Technology, Pasadena, CA 91125 USA}
\affiliation{Institute for Quantum Information and Matter, California Institute of Technology, Pasadena, CA 91125 USA}
\affiliation{Google Quantum AI, Venice, CA 90291, USA}

\author{Bryce Kobrin}
\thanks{These authors contributed equally to this work.}
\affiliation{Google Quantum AI, Venice, CA 90291, USA}
\affiliation{Department of Physics, University of California Berkeley, Berkeley, CA 94720, USA}

\author{Vincent P. Su}
\affiliation{Center for Theoretical Physics, Department of Physics, University of California, Berkeley, CA 94720, USA}

\author{Hugo Marrochio}
\affiliation{Center for Theoretical Physics, Department of Physics, University of California, Berkeley, CA 94720, USA}

\author{Norman Y.~Yao}
\affiliation{Department of Physics, Harvard University, Cambridge, MA 02138, USA}

\begin{abstract}
We analyze a simple and efficient experimental protocol to cool the Sachdev-Ye-Kitaev (SYK) model to low temperatures. The protocol utilizes local couplings between two copies of an SYK model to create a gapped adiabatic path, between a high temperature product state and a low temperature thermofield double state. By smoothly varying the coupling strength  between these two limits, one efficiently cools the SYK model. We support these predictions---and demonstrate fast cooling to the low-temperature gravitational regime---via exact numerical solutions to the large-$N$ equations of motion that govern the ground state and dynamical properties of the coupled system. Finally, we present a theoretical framework  based upon eigenstate thermalization that provides a microscopic explanation for the efficacy of the cooling protocol; intriguingly, this suggests that the protocol may be applicable for cooling strongly-interacting quantum Hamiltonians more broadly. 
\end{abstract}

\maketitle

Quantum simulators hold the potential to realize complex Hamiltonians beyond those ordinarily found in nature~\cite{altman2021quantum}. 
Among the most enticing are quantum Hamiltonians with holographic dualities to quantum gravity~\cite{hooft1993dimensional,susskind1995world,maldacena1999large}.
Exemplified by the Sachdev-Ye-Kitaev (SYK) model~\cite{sachdev1993gapless,kitaev_simple_2015,maldacena2016remarks}, these systems have already yielded numerous theoretical breakthoughs at the intersection of gravity and quantum mechanics~\cite{hayden2007black,sekino2008fast,maldacena2016bound,gao2017traversable,maldacena2017diving,yoshida2017efficient}. 
This progress has sparked a tremendous effort to realize such systems in quantum computing and simulation experiments~\cite{wei2021optical,brzezinska2023engineering,qi2019measuring,bentsen2019treelike,cao2020building,danshita2017creating,pikulin2017black,luo2019quantum,periwal2021programmable,cowsik2023engineering,gao2019traversable,brown2023quantum,nezami2023quantum,schuster2022many,jafferis2022traversable,kobrin2023comment,shapoval2023towards,landsman2019verified,blok2020quantum,sahay2024emergent,xu2020sparse,caceres2021sparse,orman2024quantum}.

Despite this interest, several hurdles remain to realizing genuine quantum simulations of quantum gravity. 
One of the most prominent is how to \emph{cool} quantum Hamiltonians with gravity duals to low temperatures, where the duality applies.
Indeed, owing to the absence of a thermal bath, reaching  low-temperature many-body states represents a broad challenge in the field of quantum simulation~\cite{terhal2000problem,cotler2019quantum,bharti2022noisy,chen2023quantum,chen2024local,kendrick2025pseudogap,lloyd2024quasiparticle,mi2024stable,farhi2000quantum,albash2018adiabatic,greiner2002quantum,simon2011quantum,semeghini2021probing,scholl2021quantum}. 
This difficulty is especially acute for Hamiltonians such as the SYK model for two reasons: ($i$) their strong non-local interactions prevent common approaches based on  quasiparticles~\cite{lloyd2024quasiparticle,mi2024stable}, simple adiabatic paths~\cite{farhi2000quantum,albash2018adiabatic,greiner2002quantum,simon2011quantum,semeghini2021probing,scholl2021quantum}, or geometric locality~\cite{hastings2005quasiadiabatic}, and ($ii$) in many applications, one desires not only low temperature thermal states, but also their purification, the so-called \emph{thermofield double} (TFD) state~\cite{maldacena2003eternal,lin2019symmetries,brown2023quantum,nezami2023quantum,schuster2022many,jafferis2022traversable,kobrin2023comment,shapoval2023towards,landsman2019verified,blok2020quantum}.
As a result, practical experimental protocols for cooling the SYK model are almost entirely lacking.
Existing approaches either require an extensive thermal bath~\cite{almheiri2019universal,zhang2019evaporation,maldacena2021syk}, large circuit depth overheads~\cite{chen2023quantum,chen2023efficient}, or variational algorithms~\cite{wu2019variational,martyn2019product,su2021variational,zhu2020generation,hastings2022optimizing} whose efficacy at large system sizes is highly unclear~\cite{mcclean2018barren}.

\begin{figure}[t]
\includegraphics[width=0.87\linewidth]{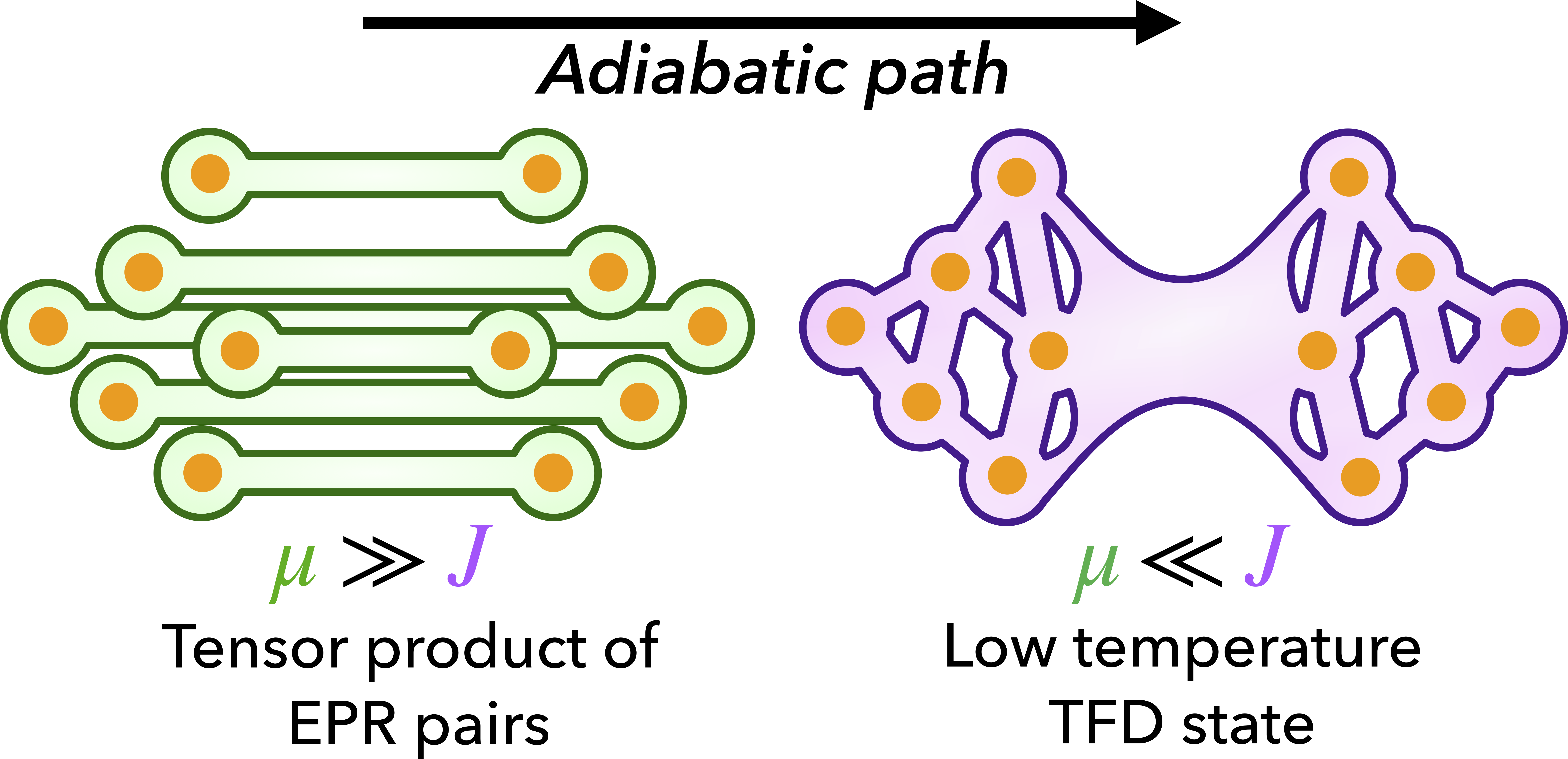}
\caption{Schematic of the experimental protocol for cooling the Sachdev-Ye-Kitaev (SYK) model~\cite{kobrin2023applications,bentsen2024approximate}. Two copies of the model, with interaction strength $J$, are coupled by pairwise terms of strength $\mu$ [Eq.~(\ref{eq:H_MQ})]. 
When $\mu \gg J$, the ground state is a tensor product of EPR pairs between the two copies (green).
When $\mu \ll J$, the ground state is a thermofield double (TFD) state at a low temperature set by $\mu$ (purple).
Our protocol prepares EPR pairs and then adiabatically interpolates to the TFD state by slowly decreasing $\mu$. 
}
\label{fig: 1}
\end{figure}

In this Letter, we propose and analyze a simple and efficient \emph{adiabatic} protocol for cooling the SYK model to low temperatures~\cite{kobrin2023applications,bentsen2024approximate}.
The protocol is based on the Maldacena-Qi Hamiltonian~\cite{maldacena2018eternal}, which couples two identical copies ($L$ and $R$) of an SYK model with pairwise two-body terms,
\vspace{-2mm}
\begin{equation} \label{eq:H_MQ}
H_{\text{MQ}} = H^{\textrm{SYK}}_L + H^{\textrm{SYK}}_R + i \mu \sum_{j=1}^N \chi_L^j \chi_R^j,
\vspace{-1mm}
\end{equation}
where $ \chi^j_{L}$, $ \chi^j_{R}$ are the Majorana fermions composing the left and right SYK models, respectively. 
The intuition behind the adiabatic protocol is as follows~\cite{bentsen2024approximate}.
When the coupling strength $\mu$ is large, the ground state of $H_{\text{MQ}}$ is a tensor product of EPR pairs between the two copies.
When $\mu$ is small, the ground state is a TFD state of the SYK model at a low temperature controlled by $\mu$~\cite{maldacena2018eternal}.
Thus, the cooling protocol simply interpolates between these two limits~\cite{kobrin2023applications,bentsen2024approximate}: One first prepares the EPR state~\footnote{We note that such a tensor product of EPR pairs can easily be prepared via a single layer of two-qubit gates or interactions.} and then adiabatically decreases $\mu$ (Fig.~\ref{fig: 1}).
Although conceptually straightforward, the efficacy of such an adiabatic cooling protocol, as well as its fundamental scalings with both system size and desired temperature, remain essential open questions~\cite{supp}.

To this end, our main results are fourfold.
First, we perform large-$N$ Schwinger-Dyson numerical simulations, which verify that the coupled Hamiltonian is gapped, with a ground state resembling the TFD state, at all values of $\mu$.
From the adiabatic theorem, this enables cooling to low temperatures, $1/\beta$, in time, $\mathcal{O}( N J^{-1} \log( \beta J ) )$, where $N$ is the number of fermions and $J$ is the SYK interaction strength.
Second, we show that one can cool much faster than this, in times $\mathcal{O}( \beta \log( \beta J) )$, by exploiting \emph{semi-classical} dynamics in the excited states  of the coupled Hamiltonian.
To support this, we directly simulate our  protocol at large $N$ and observe rapid cooling to temperatures as low as $(\beta J)^{-1} \approx 1/40$ in short times. 
Third, building on earlier work~\cite{cottrell2019build}, we provide a heuristic  framework based on  eigenstate thermalization~\cite{deutsch2018eigenstate} to analyze the behavior of the coupled Hamiltonian at intermediary $\mu$~\cite{fn10}, which closely matches our numerical results. 
Finally, we perform state-vector numerical simulations, which show that our protocol enables efficient cooling even in finite-size systems outside of the large-$N$ regime.

\emph{The coupled Hamiltonian.}---We begin with a brief review of the coupled Hamiltonian, Eq.~(\ref{eq:H_MQ})~\cite{maldacena2018eternal,zhou2020tunneling,qi2020coupled,lensky2021rescuing,maldacena2021syk,milekhin2024measurement}.
The Hamiltonian acts on a system of $2N$ Majorana fermions, $\{ \chi^j_{L}, \chi^j_{R} \}_{j=1}^N$, on sides $L$ and $R$, where $\{\chi^i_L, \chi^j_L\} = \delta_{ij}$.
The individual SYK models are, $H_{\alpha} = \sum_{ijkl} J_{ijkl} \chi^i_\alpha \chi^j_\alpha \chi^k_\alpha \chi^l_\alpha$, for $\alpha = L,R$, where $J_{ijkl}$ are Gaussian random variables with  variance $6 J^2/N^3$~\cite{maldacena2016remarks}.
Our goal is to utilize the coupled Hamiltonian to prepare low temperature TFD states, $\sum_m e^{-\beta E_m/2} \ket{m} \otimes \ket{m^*} / \sqrt{\mathcal{Z}}$, of the individual SYK models.
Here, $\ket{m}$ are the SYK eigenstates, with energies $E_m$, and $\mathcal{Z} = \sum_m e^{-\beta E_m}$.
A thermal state can be obtained from the TFD state by tracing out $L$ or $R$.

At large $\mu$, the ground state of Eq.~(\ref{eq:H_MQ}) is a tensor product of EPR pairs between the two sides.
This follows because each coupling, $i \chi_L^j \chi_R^j$, projects onto an individual EPR pair between the fermions $\chi_L^j$ and $\chi_R^j$~\cite{fn1}.

At small $\mu$, the ground state is a TFD state at a low temperature set by $\mu$~\cite{maldacena2018eternal}. 
The  low-energy dynamics are governed by a semi-classical gravity dual, analogous to that of the individual SYK models~\cite{kitaev_simple_2015,maldacena2018eternal}.
In the gravitational picture, the TFD state corresponds to a wormhole of length $\beta/4 \pi$ between the two sides.
Excitations above the ground state are gapped, $\mathcal{O}(\beta^{-1})$, and come in two distinct classes.
The first are ``matter'' excitations.
These are obtained e.g.~by applying a local fermion operator to the ground state, $\chi^j_L \ket{ \psi_{\text{gs}} }$, and correspond to massive particles that move on the gravitational geometry~\cite{maldacena2018eternal}.
The second are ``graviton'' excitations.
These are obtained e.g.~by initializing the system in a TFD state at an altered value of $\beta$, and correspond to fluctuations in the spacetime geometry itself~\cite{maldacena2018eternal}.

The behavior of the coupled Hamiltonian beyond these two limits is less explored.
Nonetheless, a heuristic argument suggests that the coupled ground state might smoothly interpolate between the EPR state and the TFD state as we lower $\mu$.
The  Hamiltonian features a competition between two terms: the individual SYK models and the couplings.
The former favor a low energy (with respect to $H_{L,R}$) on each side.
The latter favor entanglement between the two sides; by monogamy of entanglement, this requires a large \emph{entropy} on each side.
Thus, much like a system in a thermal bath, the ground state is determined by a competition between energy and entropy, with the coupling strength $\mu$ playing, roughly, the role of the temperature.
In what follows, we will show that this simple intuition is remarkably accurate, which enables efficient cooling to low temperatures.

\begin{figure}[t]
\centering
\includegraphics[width=\linewidth]{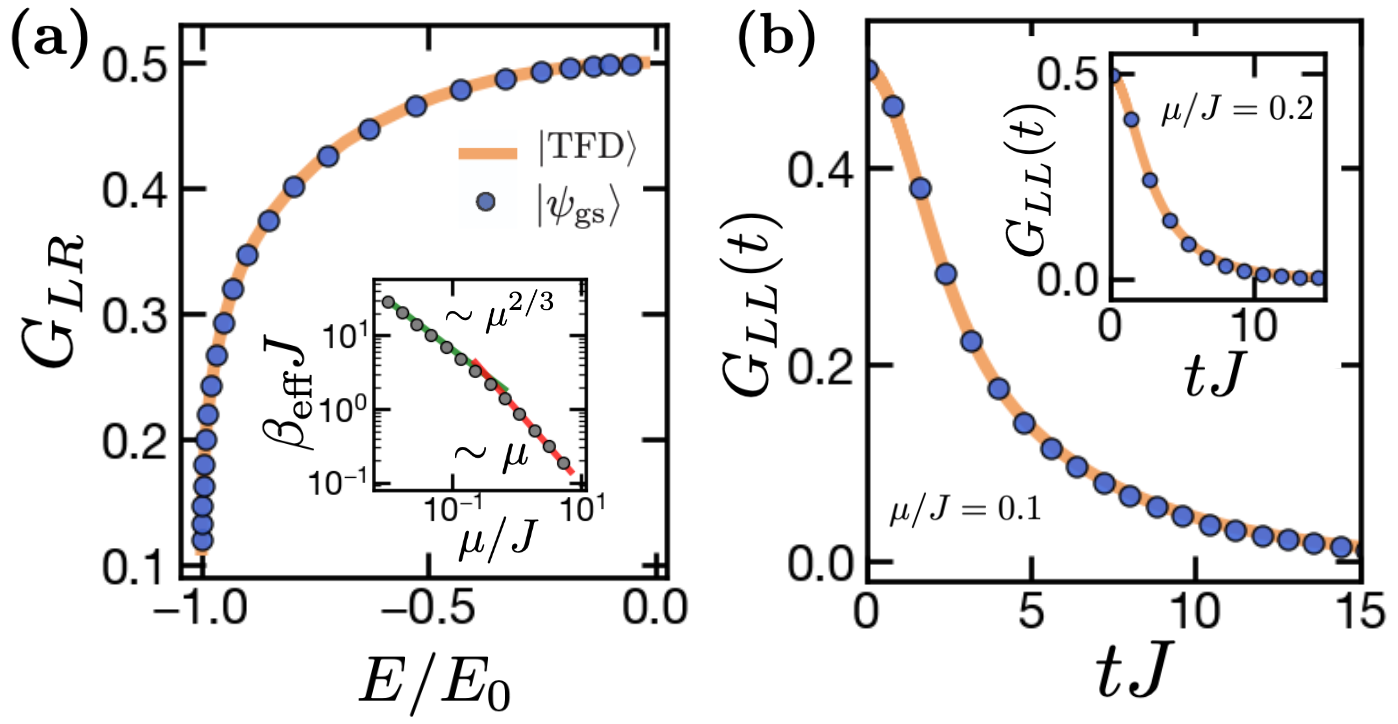}
\caption{Numerical comparison of the ground state of Eq.~(\ref{eq:H_MQ}) (blue) and the TFD state (orange) in the large-$N$ limit. \textbf{(a)} The two-side correlator, $G_{LR} \equiv \left<i\chi_L \chi_R \right>$,  vs. the single-side energy, $E \equiv \left <H_L\right>$, normalized by the magnitude of the ground-state energy of the SYK model, $E_0$. 
Inset: Matching the value of $G_{LR}$ yields the effective TFD temperature $\beta_{\textrm{eff}}$ as a function of $\mu$. 
\textbf{(b)} The single-side auto-correlation function, $G_{LL}(t) \equiv \left<i\chi_L(t) \chi_L(0) \right>$, for the ground state of Eq.~(\ref{eq:H_MQ}) at $\mu = 0.1 J$ and $\mu = 0.2 J$ (inset), and the TFD state at $\beta J = 3.7 $ and $\beta J = 6.1 $, respectively, which correspond to $\beta \equiv \beta_{\text{eff}}(\mu)$.} 
\label{fig: ground state}
\end{figure}

\begin{figure*}[t]
\centering
\includegraphics[width=\linewidth]{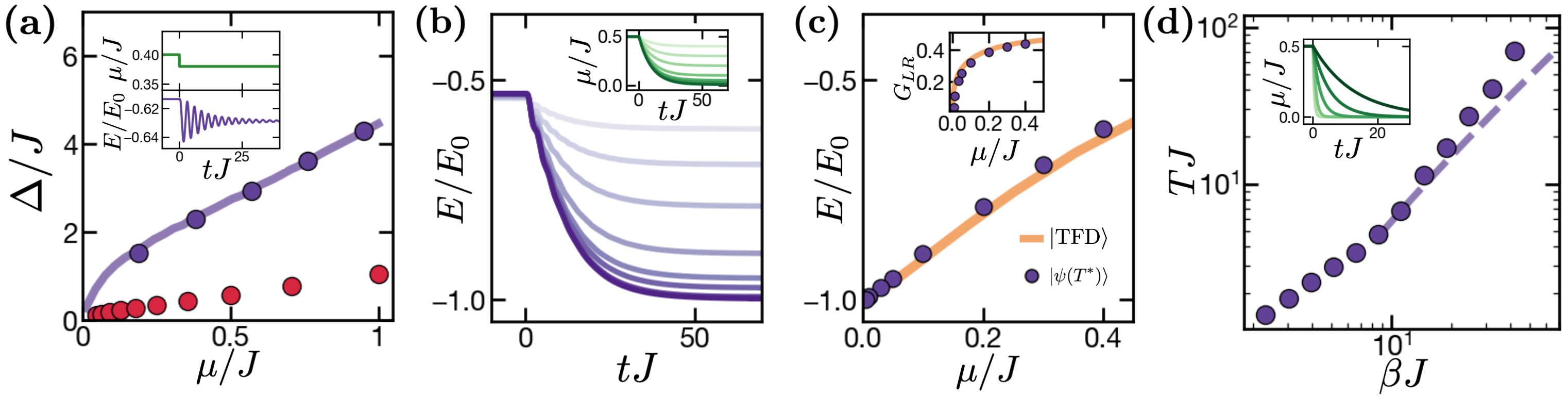}
\caption{Dynamics of the coupled Hamiltonian in the large-$N$ limit.
In all cases, the initial state is a low-temperature state of Eq.~(\ref{eq:H_MQ}) at an initial coupling strength $\mu_i$. 
\textbf{(a)} Quench dynamics upon a small sudden change in the coupling strength. 
The change leads to oscillations in the single-side energy, $E \equiv \left <H_L\right>$.
Inset: Example for $\mu_i = 0.4 J$ and $\mu_f = 0.38 J$.
The oscillation frequency (purple circles) closely matches the graviton gap predicted by the Schwinger-Dyson equations and our thermalization-based framework (purple curve) across a range of $\mu \equiv \mu_i$~\cite{supp}.
For comparison, the matter gap (red) of Eq.~(\ref{eq:H_MQ}), obtained by simulating Eq.~(\ref{eq:H_MQ}) in imaginary time~\cite{garcia2019quantum}, is moderately smaller. 
\textbf{(b)} Simulation of our proposed adiabatic protocol.
The coupling strength is decreased adiabatically from $\mu_i = 0.5 J$ to various $\mu_f$, as $\mu(t) = (\mu_i-\mu_f) e^{-t/T_*}+\mu_f$ with $T_* = 10/J$ (inset). 
The single-side energy (purple) is observed to decrease alongside $\mu(t)$ and saturates a final value determined by $\mu_f$. 
\textbf{(c)} Comparison between the final value of the single-side energy (purple circles) and its value in the TFD state at $\mu_f$ (orange curve).
The small offset at large $\mu_f$ arises from the small  initial physical temperature used to regulate the large-$N$ numerical simulations~\cite{supp}. Inset: Identical comparison for the two-sided correlator, $G_{LR}$. \textbf{(d)} Modified adiabatic protocol with $\mu_f = 0$, i.e.~$\mu(t) = \mu_i \exp(-t/T_*)$ (inset). We sweep $T_*$ to determine the minimum time $T$ required to reach a  single-side energy corresponding to a desired temperature $\beta^{-1}$ (purple circles). The theory prediction $T = a \beta \log (\beta)$ is shown for $a = 0.25$ (dashed).}
\label{fig: adiabatic}
\end{figure*}

\emph{Ground state and gapped adiabatic path.}---We begin by characterizing the ground state and gap of the coupled Hamiltonian. 
At large system sizes $N$, the dynamics are  governed by the Schwinger-Dyson equations. 
Under imaginary time-evolution, these are given by ~\cite{maldacena2018eternal,fn2}
\begin{equation}
\begin{aligned}
\partial_{\tau} \textbf{G}(\tau) &= (\bm{\Sigma}(\tau) - i \bm{\mu}(\tau)) * \textbf{G}(\tau)+\bm{\delta}(\tau) \\
\bm{\Sigma}(\tau) &= J^2\textbf{G}^3(\tau)
\end{aligned}
\end{equation}
where $\textbf{G}(\tau)$ and $\bm{\Sigma}(\tau)$ are auxiliary matrix variables related to local fermion correlation functions on the left and right sides, $*$ denotes convolution under imaginary time-evolution, $\bm{\mu}(\tau)$ is specified by the time-dependent coupling, and $\bm{\delta}(\tau)$ is the Kronecker delta function; see the supplemental material (SM) for  details~\cite{supp}.
The equations can be numerically solved in both imaginary and real time, yielding ground state and dynamical properties.

We evaluate the closeness of the ground state and TFD state in two ways.
First, we compare the expectation value of the coupling operator, $G_{LR} \equiv \left <i\chi_L \chi_R \right >$, as a function of the single-side energy density, $\left < H_L \right >$, in each state [Fig.~\ref{fig: ground state}(a)].
Encouragingly, we observe near perfect agreement at all $\mu$.
From this agreement, we can  extract the  temperature $\beta^{-1}$ corresponding to each value of $\mu$. 
We observe two distinct regimes, consistent with expectations: at large $\mu$, $\beta^{-1} \! \sim \! \mu$, and at small $\mu$, $\beta^{-1} \! \sim \! J^{1/3} \mu^{2/3}$~\cite{maldacena2018eternal}.

Second, we compare the dynamics of local auto-correlation functions, $G_{LL}(t) \equiv \left <i \chi_L(t) \chi_L (0) \right >$, when either the ground or TFD state are quenched to evolve under the individual uncoupled SYK models.
Setting $\beta$ according to $\mu$ as above, we observe extremely close agreement between the two dynamics for all  $\mu$ [Fig.~\ref{fig: ground state}(b)].

Having established the resemblance of the ground state and TFD state, we now address the spectral gap.
To be specific, the Schwinger-Dyson equations can measure the gap with respect to \emph{local} perturbations in the large-$N$ limit.
This is achieved by Fourier transforming local correlation functions as a function of time, and isolating the minimum non-zero frequency.

Motivated by the aforementioned two classes of excitations at small $\mu$, we analyze two local correlation functions of the coupled Hamiltonian.
First, we consider $G^\mu_{LL}(\tau) \equiv \left <i \chi_L(i \tau) \chi_L (0) \right >$, where in contrast to Fig.~\ref{fig: ground state}(b), we evolve under imaginary time and the coupling remains present during time-evolution.  
When $\mu$ is small, this function diagnoses the  gap to matter excitations~\cite{maldacena2018eternal}. 
We find that the gap, $\Delta_m$, remains non-zero for all $\mu$ and  smoothly grows as $\mu$ increases [Fig.~\ref{fig: adiabatic}(a)].
Second, we consider a quenched correlation function, where the system begins in a TFD state at one value of $\mu$, and is quenched to an altered value.
When $\mu$ is small, this diagnoses the gap to graviton excitations~\cite{supp}. 
Again, we find that the gap, $\Delta_g$, remains non-zero for all $\mu$ and smoothly grows as $\mu$ increases [Fig.~\ref{fig: adiabatic}(a)].

From these observations, the  adiabatic theorem~\cite{griffiths2018introduction} immediately implies that our protocol can efficiently prepare low temperature TFD states by interpolating from large to small $\mu$.
We proceed as in Fig.~\ref{fig: 1}, and consider an exponential profile, $\mu(t) = \mu_i (\mu_f/\mu_i)^{(t/T)}$, over time. 
This ensures that the Hamiltonian changes more slowly as $\mu$ decreases, where the gap becomes smaller.
To estimate the requisite protocol time $T$, we perform a standard adiabatic analysis.
We find that the leading many-body infidelity arises from leakage to matter excited states, with a magnitude estimated by Fermi's golden rule~\cite{supp}, 
\begin{equation} \label{eq: leakage main text}
\varepsilon \approx \int_0^T dt \, \sum_{j=1}^N \frac{| \langle \psi_{m,j} | \partial_t | \psi_{\text{gs}} \rangle |^2}{\Delta_m} = \mathcal{O}\left( \frac{N  \log(\beta J)}{JT} \right),
\end{equation}
where $\ket{\psi_{m,j}}$ denotes a matter excited state corresponding to fermion $j$.
The infidelity grows linearly in the system size $N$, owing to the sum over excited states.
This yields a protocol time, $T = \mathcal{O}(N J^{-1} \log(\beta J) )$, which also scales linearly in $N$.

\emph{Fast cooling.}---The above linear scaling in the system size $N$ arises because the standard adiabatic approach guarantees a high \emph{many-body} fidelity between the prepared state and the true ground state~\cite{fn9}.
We will now show that much faster cooling can be achieved, in times independent of the system size $N$, if one only desires close agreement in local observables and correlation functions.
Crucially, such an agreement is sufficient for most experimental proposals involving the SYK model~\cite{fn11}.

To address this fast cooling regime, we must analyze the excited state dynamics of the coupled Hamiltonian.
This is necessary because, following our earlier analysis, the prepared state will have a negligibly small many-body fidelity with the true ground state if the protocol time is sublinear in $N$.
Fortunately, the excited state dynamics of Eq.~(\ref{eq:H_MQ}) take an exceptionally simple and known form~\cite{maldacena2018eternal}. 
In particular, at small $\mu$, the graviton excitations are governed by an effective \emph{one-dimensional} Hamiltonian~\cite{maldacena2018eternal,supp},
\begin{equation} \label{eq: eff H main text}
	\frac{H_{\text{eff}}}{JN} =  \frac{p^2}{2 m} + V(\phi;\mu), \,\,\,  V(\phi; \mu) = e^{2\phi} - a \mu e^{-2\Delta \phi},
\end{equation}
with respect to a coordinate $\phi$, its conjugate  $p$, and a potential $V(\phi;\mu)$; here, $a = \mathcal{O}(J^{-1})$ and $m = \mathcal{O}(1)$.
The ground state of $H_{\text{eff}}$ corresponds to the ground state of Eq.~(\ref{eq:H_MQ}), and the excited states correspond to graviton excitations~\cite{maldacena2018eternal}.

Physically, $H_\text{eff}$ acts on a subspace of thermofield double-like many-body states, where the coordinate $\phi$ controls the effective TFD temperature~\cite{maldacena2018eternal,supp}.
The ground state of $H_\text{eff}$ is concentrated at the minimum of the potential $V(\phi;\mu)$; the location of the minimum, $\phi^*(\mu)$, determines the temperature of the corresponding TFD state.
The low-lying excited states of $H_\text{eff}$ are captured by Taylor expanding $V(\phi;\mu)$ about its minimum in a harmonic oscillator fashion.
Crucially, owing to the normalization factor of $N$, $H_{\text{eff}}$ also features a \emph{semi-classical} limit controlled by an effective Planck's constant, $1/N$.
In this limit, the low-lying excited state dynamics are simply given by the classical equations of motion for a harmonic oscillator centered at $\phi^*(\mu)$.

This simple description leads to the following physical picture for fast cooling.
At time zero, the system is initialized at the minimum, $\phi^*(\mu_i)$, corresponding to a high effective temperature controlled by $\mu_i$. 
Over time, the value of $\mu$ is lowered, and the location of the harmonic oscillator minimum slowly interpolates to lower effective temperatures.
If the interpolation is sufficiently slow, the semi-classical dynamics will remain nearby the minimum for all times, resulting in the preparation of a state near the final minimum, $\phi^*(\mu_f)$.
This corresponds to a TFD-like state at a low effective temperature controlled by $\mu_f$.
The state has a low many-body overlap with the true ground state, owing to small order-one fluctuations of the effective temperature about the potential minimum, but a good agreement in local observables.

To precisely characterize this fast-cooling regime, we perform a thorough analysis of errors arising from the protocol above (see 
the supplemental material for full details~\cite{supp}).
We find that two sources of error contribute at leading order.
First, as aforementioned, the motion of the  minimum during the cooling protocol leads to small fluctuations, $\delta \phi$, about the minimum in the final state.
We estimate that these lead to effective temperature fluctuations, $\delta \beta / \beta = \mathcal{O}(\log^2(\beta J)/\beta^{-2} T^2)$~\cite{supp}.
Second,  the protocol generates a finite density of \emph{matter} excitations, $n_m \approx \varepsilon / N$, where $\varepsilon$ is given by Eq.~(\ref{eq: leakage main text}).
Each excitation carries an energy of order the gap, $\Delta_m \sim \beta^{-1}$; this leads to an effective temperature fluctuations, $\delta \beta / \beta = \mathcal{O}(\log(\beta J) / \beta^{-1} T)$~\cite{supp}.
Both of these effects contribute to an exponentially small many-body fidelity, but yield good agreement in local observables when $\delta \beta / \beta$ is small~\cite{supp}.
This requires a protocol time $T = \mathcal{O}(\beta \log(\beta J))$.

To test these predictions, we numerically simulate the dynamics of the fast-cooling protocol using the large-$N$ Schwinger-Dyson equations. 
Indeed, the fast-cooling regime is the \emph{only} limit of the cooling protocol that can be directly simulated using the large-$N$ Schwinger-Dyson equations, since the naive adiabatic approach requires time scaling as $N \rightarrow \infty$.
For technical reasons~\cite{supp}, the large-$N$ simulations cannot be performed at infinite $\mu$, so we  initialize the system in a high-temperature TFD state at a large but finite $\mu_i = 0.5 J$ instead~\cite{fn_large_N}.
In Fig.~\ref{fig: adiabatic}(b) we plot the single-side energy as a function of time as the coupling $\mu(t)$ is gradually lowered from $\mu_i$ to various final values $\mu_f$.
We observe that the energy smoothly decreases as a function of time to a final value that closely agrees with the TFD prediction at $\mu_f$ [Fig.~\ref{fig: adiabatic}(c)].
We also observe that the final value of the two-side correlator, $G_{LR}$, agrees with the TFD prediction [Fig.~\ref{fig: adiabatic}(c) inset].
This confirms our prediction that the adiabatic protocol can cool to low-temperature TFD states (to within good agreement in local correlation functions) in a time independent of the system size.

To determine the scaling of the required protocol time with the desired temperature $\beta^{-1}$, we consider a modified adiabatic protocol, in which the coupling exponentially interpolates between $\mu_i = 0.5 J$ and $\mu_f = 0$~\cite{fn8}.
Under this protocol, we expect the
system to remain close to a TFD state down to a threshold temperature set by the interpolation speed, after which non-adiabatic effects halt further cooling. 
By sweeping the interpolation speed, we determine the required time $T$ to cool to various temperatures $\beta^{-1}$ [Fig.~\ref{fig: adiabatic}(d)].
We observe a gradual increase in the required protocol time with the desired temperature; for $\beta \gtrsim 10$, our numerical results approximately agree with our theory prediction, $T \sim \beta \log(\beta)$.

\emph{Thermalization-based  framework.}---Thus far, our study of the intermediate-$\mu$ regime has been entirely numerical.
We now introduce a complementary heuristic theoretical framework to understand the coupled Hamiltonian for general $\mu$.
Our approach builds on earlier work~\cite{cottrell2019build}, but features substantially improved error estimates and the first explicit application to the SYK model.

\begin{figure}[t]
\centering
\includegraphics[width=0.95\linewidth]{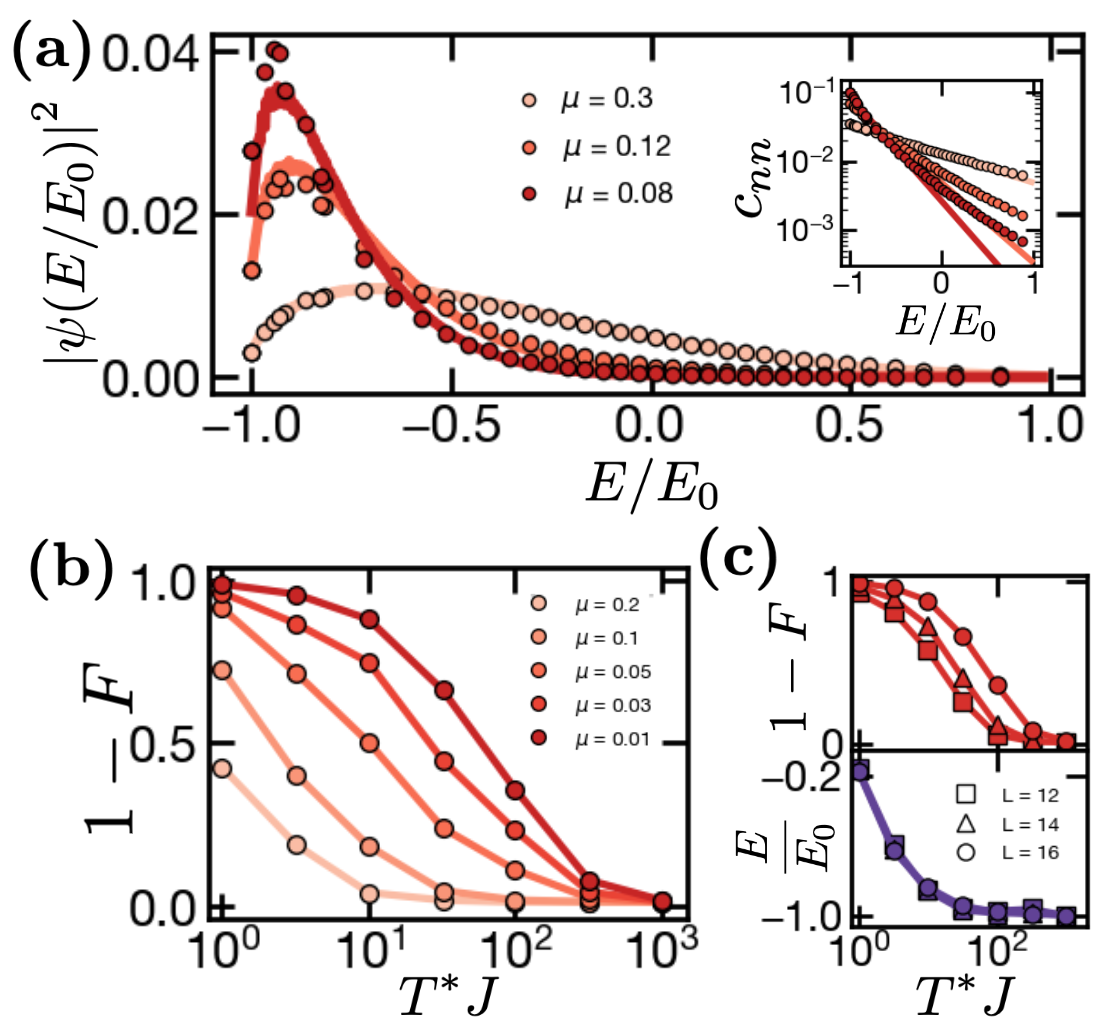}\caption{Finite-size numerical analysis of the ground state and cooling dynamics of the coupled Hamiltonian. \textbf{(a)} The re-scaled coefficients, $\psi_{nn} \equiv \sqrt{D_n} c_{nn}$, of the ground-state wavefunction (circles) and the TFD state (curves) for various $\mu$ ($2N = 48$). We plot only a subset of coefficients for visual clarity. Inset: The bare coefficients $c_{nn}$ plotted as in Ref.~\cite{garcia2019quantum}. \textbf{(b)} Dynamical simulation of our adiabatic protocol ($2N = 32$). The system begins in the EPR state and the coupling decreases as $\mu(t) = \mu_i (\mu_f/\mu_i)^{(t/T_*)}$ with $\mu_i = 4J$. 
The infidelity of the final state with the ground state of Eq.~(\ref{eq:H_MQ}) is shown for various $T_*$ and $\mu_f$.
\textbf{(c)} The infidelity at $\mu_f = 0.01 J$ for various system sizes $N = 12,14,16$ (top). The  single-sided energy at $\mu_f = 0.01 J$ for the same system sizes (bottom).}
\label{fig: ED}
\end{figure}

We consider the matrix elements of the coupled Hamiltonian in the energy eigenstates, $\ket{m} \otimes \ket{n^*} \equiv \ket{mn^*}$, of the individual SYK models~\cite{cottrell2019build}.
Notably, the TFD state lies within the subspace of ``paired'' eigenstates, $m = n$.
Motivated by this,  our approach centers on a single simplifying assumption: that the physics of Eq.~(\ref{eq:H_MQ}) can be captured solely through its action  on the paired subspace.

Within the paired subspace, the coupled Hamiltonian takes a simple form. 
The individual SYK models contribute a linear potential energy, $\sum_n 2E_n \dyad{nn}$.
The couplings contribute hoppings between paired states, $\mu \bra{n n^*} i \chi_L^j \chi_R^j \ket{m m^*} = \mu | \! \bra{n} \!\chi^j\! \ket{m} \! |^2$, whose magnitudes are proportional to off-diagonal matrix elements of the local fermion operators.
To proceed, we invoke the eigenstate thermalization hypothesis (ETH)~\cite{deutsch2018eigenstate}, which posits that $| \! \bra{n} \!\chi^j\! \ket{m} \! |^2$ is a smooth function of $E_n$, $E_m$ with support in a narrow range $|E_n-E_m| \lesssim \beta^{-1}$~\cite{supp}.
We can then take a continuum limit, which yields an effective one-dimensional Schrodinger equation~\cite{cottrell2019build,fn12,supp},
\begin{equation} \label{eq: main text SHO}
 i N^{-1} \partial_t \psi(e) = [2e - \mu G(e) ] \psi(e) - \mu G^{''}\!(e) \frac{\partial_e^2}{2 N^2} \psi(e),
\vspace{-1mm}
\end{equation}
where $e \equiv E/N$, $G(e) \equiv \langle \chi^j \sqrt{\rho_\beta} \chi^j \sqrt{\rho_\beta} \rangle$ is the two-point function at $\beta$ corresponding to $e$, $\sum_n \frac{\psi(e_n)}{\sqrt{D_n}} \ket{nn}$ is the associated many-body state, and $D_n$ the density of states. 

Intriguingly, we find that this effective Schrodinger equation yields surprisingly accurate predictions for the ground state and graviton dynamics of the coupled Hamiltonian.
We confirm this in three ways.
First, when $\mu$ is small, we find that Eq.~(\ref{eq: main text SHO}) exactly reproduces Eq.~(\ref{eq: eff H main text}) upon a change of coordinates~\cite{supp}.
Second, we analyze a basic self-consistency check: whether the variance of the \emph{full} Hamiltonian in the \emph{paired-subspace} ground state is small~\cite{fn6}.
A basic computation yields a dimensionless variance $\mathcal{O}(\mu^2 \beta^2 (1-G^2_\beta))$, which is small at both small $\mu$, where $\mu \beta \ll 1$, and large $\mu$, where $G_\beta \approx 1$.
This supports the use of our framework in both the high and low temperature regimes.
Finally, to address the intermediate regime, we benchmark Eq.~(\ref{eq: main text SHO}) against our large-$N$ numerics.
In Fig.~\ref{fig: ground state}(a), we observe that Eq.~(\ref{eq: main text SHO}) provides accurate predictions for the effective temperature of the ground state at all $\mu$.
In addition, in Fig.~\ref{fig: adiabatic}(a), we observe that Eq.~(\ref{eq: main text SHO}) provides similarly accurate predictions for the graviton oscillation gap at all $\mu$~\cite{fn5}.

\emph{Finite-size numerics.}---We conclude by presenting numerical studies at finite sizes. 
Utilizing massively parallel, matrix-free Krylov subspace methods, we first compute the ground-state of the coupled system with $2N=48$ Majoranas.  
Following our framework in the previous section, in Fig.~\ref{fig: ED}(a) we plot the re-scaled coefficients $\psi_{nn} \equiv \sqrt{D_n} c_{nn}$ within the paired subspace, where $D_n$ is determined analytically~\cite{garcia2017analytical}.
We observe good agreement with TFD predictions, consistent with prior observations that the ground and TFD states have high many-body overlap at small sizes~\cite{maldacena2018eternal,garcia2019quantum}.
As in earlier work, we also observe disagreement in the individual coefficients between the ground and TFD state  at high energies where the coefficients are already negligibly small [Fig.~\ref{fig: ED}(a) inset] \cite{garcia2019quantum}.
We remark that our thermalization-based framework suggests that one should not expect an exact agreement between the two wavefunctions.

We next use our finite-size numerics to directly simulate the cooling protocol. 
This complements our large-$N$ studies by allowing access to the many-body overlap of the prepared and TFD states, and by allowing larger initial values of $\mu_i$. 
We observe that the required protocol time increases gradually as $\mu_f$ decreases, as expected [Fig.~\ref{fig: ED}(b)].
The required time to obtain a high many-body overlap also increases with the system size [Fig.~\ref{fig: ED}(c) top], consistent with our adiabatic analysis. 
Finally, the required time to achieve a small single-side energy density, instead of a small many-body fidelity, is independent of system size [Fig.~\ref{fig: ED}(c) bottom], as expected from our large-$N$ studies and theoretical analysis.

\emph{Outlook.}---Our work demonstrates that one of the primary hurdles to realizing quantum gravity in the lab is in fact imminently achievable in quantum experiments.
This raises several intriguing questions for future work. 
First, does this cooling approach extend to recent \emph{bosonic} models of holography~\cite{swingle2024bosonic,liu2024signature}, which may be significantly easier to experimentally realize?
Second, our results imply upper bounds on the  complexity of low temperature SYK states~\cite{fn13}.
Do these bounds bear any connection to the conjectured complexity-action correspondence~\cite{brown2016holographic}?
Finally, this approach provides a novel method to cool quantum systems, by leveraging entanglement with an auxiliary copy. 
Can this approach extend to more general strongly-interacting quantum Hamiltonians?

\textit{Acknowledgements}---We are grateful to Eric Anschuetz, Chi-Fang Chen, Robbie King, Xiao-Liang Qi, and Zhenbin Yang for discussions and insights.
We are especially grateful to Alexey Milekhin for valuable advice on our large-$N$ numerical simulations and several discussions.
For the finite-size numerics, we utilize the \texttt{dynamite} Python frontend \cite{dynamite}, which supports a matrix-free implementation of Krylov subspace methods based on the \texttt{PETSc} and \texttt{SLEPc} packages.
This work was supported in part by the Department of Energy under QuantISED award DE-SC0019380 and the Simon's Foundation. 
T.S. acknowledges
support from the Walter Burke Institute for Theoretical Physics at Caltech and the U.S. Department of Energy, Office of Science, National Quantum Information Science Research Centers, Quantum Systems Accelerator.
The Institute for Quantum Information and Matter is an NSF Physics Frontiers Center (NSF Grant PHY-2317110).

\let\oldaddcontentsline\addcontentsline
\renewcommand{\addcontentsline}[3]{}
\bibliography{refs}

\begin{thebibliography}{112}
\bibitem [{fn1({\natexlab{a}})}]{fn10}%
  \BibitemOpen
  \href@noop {} {\bibfield  {journal} {\bibinfo  {journal} {Here, by intermediate $\mu$, we mean any $\mu$ that is not much greater nor much less than one}\ } ({\natexlab{a}})}\BibitemShut {NoStop}%
\bibitem [{fn2()}]{fn2}%
  \BibitemOpen
  \href@noop {} {\bibinfo  {journal} {We note that, to be precise, the Schwinger-Dyson equations apply only for $J, \mu = \mathcal{O}(1)$ and so do not apply to the strict $\mu \rightarrow \infty$ limit in which the ground state is equal to EPR pairs. Thus, our numerical simulations begin in the ground state at some finite value of $\mu$ (i.e. a large-but-not-infinite temperature TFD state), and then cool to arbitrarily low temperatures. Our theoretical framework suggests that the same physics extend to $\mu \rightarrow \infty$ without modification}\ }\BibitemShut {NoStop}%
\bibitem [{fn8()}]{fn8}%
  \BibitemOpen
\bibfield  {journal} {  }\href@noop {} {\bibinfo  {journal} {This modified adiabatic protocol is chosen so that we need only sweep one parameter, $T_*$, instead of two parameters, $T_*$ and $\mu_f$, in the original adiabatic protocol}\ }\BibitemShut {NoStop}%
\bibitem [{fn5()}]{fn5}%
  \BibitemOpen
\bibfield  {journal} {  }\href@noop {} {\bibinfo  {journal} {Interestingly, we also observe a slow decay in the amplitude of these oscillations over time [Fig.~\ref{fig: adiabatic}(a) inset]. This goes beyond the physics predicted by the framework, and could arise either as a technical artifact of the non-zero physical temperature in our large-$N$ numerics, or from corrections to the paired-subspace ansatz at intermediate coupling strengths}\ }\BibitemShut {NoStop}%
\end{thebibliography}
\let\addcontentsline\oldaddcontentsline

\onecolumngrid
\newpage

{\centering
\large\bfseries
Supplementary Material: Cooling the Sachdev-Ye-Kitaev model using thermofield double states
\par}

\tableofcontents

\section{Relation to previous work} \label{app: relation}

In this section, we briefly discuss the relation between our results and Refs.~\cite{kobrin2023applications,bentsen2024approximate}, where the adiabatic cooling protocol was introduced.
Ref.~\cite{bentsen2024approximate} introduced the cooling protocol as a counter-example to the no-low-energy-trivial-state (NLTS) conjecture for the SYK model, alongside several independent results on error-correcting properties of low-temperature SYK states.
To support the success of the cooling protocol, Ref.~\cite{bentsen2024approximate} utilized large-$N$ Schwinger-Dyson numerical simulations to establish two properties: (i) the coupled Hamiltonian is gapped for all $\mu$, with a gap that scales $\sim \mu$ at large $\mu$ and as $\sim \mu^{2/3}$ at small $\mu$, and (ii) the single-side energy of the ground state of the coupled Hamiltonian can be tuned arbitrarily close to the SYK ground state energy.
The first result demonstrates the existence of a gapped adiabatic path to the low-temperature gravitational regime.
The second result shows that the low-temperature regime can reach arbitrarily close to the SYK ground state energies.
Ref.~\cite{kobrin2023applications} is a Ph.D. thesis by one of the authors of this work, whose results on the cooling protocol are incorporated into our current manuscript.

Owing to our focus on the eventual application of the cooling protocol in quantum simulation experiments, the numerical and theoretical analyses in our work are significantly more thorough and refined than those of Refs.~\cite{kobrin2023applications,bentsen2024approximate}.
Our extended analyses paint a more subtle yet cohesive picture of the success and efficiency of the cooling protocol for the SYK model.
We highlight a few of the most significant additions from our work below:
\begin{itemize}
    \item We demonstrate close agreement between the ground state and the thermofield double state in local observables and correlation functions at all values of the coupling strength $\mu$. The strong agreement at intermediate $\mu$ is particularly surprising and not anticipated in earlier works.
    
    \item More importantly, we provide detailed analyses of the scaling of the cooling protocol with the system size $N$. This reveals a subtle distinction between two different regimes of the cooling protocol: a slow regime where the time scales linearly in $N$ and guarantees a high many-body fidelity, and a fast regime where the time is independent of $N$ and guarantees good agreement in local observables. In particular, despite naive expectations, the realization of fast cooling in time independent of $N$ does not follow merely from the gap of the Hamiltonian, owing to an extensive number of couplings to excited states during the cooling protocol~\cite{fn9}. Rather, to establish fast cooling, one instead requires a more substantial analysis of  excited state dynamics, which we perform. 
    \item We also provide explicit computations of the leading-order sources of error in the cooling protocol, in both the fast and slow cooling regimes. This allows us to provide explicit predictions for the scaling of the protocol with $\beta$ in addition to the scaling with $N$.
    \item Perhaps most significantly, we substantiate these theoretical predictions by performing an extensive set of direct numerical simulations of the cooling protocol in both the large-$N$ limit and at finite system sizes.
    These numerical simulations provide the strongest possible confirmation of the success of the cooling protocol, and its scaling with $N$ and $\beta$. 
    In addition, the rapid practical speed of cooling observed in these numerics also provides support  for the use of the cooling protocol in eventual practical applications.
    \item Finally, we build on Ref.~\cite{cottrell2019build} to provide a simple theoretical framework based on eigenstate thermalization to capture many behaviors of the cooling protocol both within and above the low-temperature gravitational regime. The precise agreement of this framework with large-$N$ numerical simulations at intermediate $\mu$ is particularly non-trivial and surprising, and provides an important analytic window into the behavior of the coupled Hamiltonian in this regime.
\end{itemize}

\section{Background on the coupled Hamiltonian} \label{app: MQ}

Here, we provide a few essential details on the coupled Hamiltonian.
In Section~\ref{sec: fermion TFD}, we review the definition of the EPR and TFD state for fermionic systems.
In Section~\ref{sec: SD eqs}, we review the Schwinger-Dyson equations that govern the large-$N$ limit of the coupled Hamiltonian~\cite{maldacena2018eternal}.

\subsection{The fermionic TFD state} \label{sec: fermion TFD}

The EPR state on $2N$ fermions is defined by the condition,
\begin{equation} \label{eq: EPR condition}
( \chi^j_L + i \chi^j_R )\ket{\text{EPR}} = 0,
\end{equation}
for all $j = 1,\ldots,N$.
From the EPR state, the TFD state is defined as
\begin{equation}
\ket{\text{TFD}} = \frac{1}{\sqrt{\mathcal{Z}}} e^{-\beta(H_L+H_R)/4} \ket{\text{EPR}},
\end{equation}
where $\mathcal{Z} = \tr( e^{-\beta H} )$ is the partition function of $H$.

The EPR state can be obtained from an initial product state via a depth-1 Gaussian unitary.
For example, if one pairs the Majorana fermions $2j, 2j+1$ on either side of the system into complex fermions, and begins in the zero fermion state, $(\chi_\alpha^{2j} + i \chi_\alpha^{2j+1}) \ket{0} = 0$ for $\alpha = L,R$, then we have 
\begin{equation}
    \ket{\text{EPR}} = \prod_{j=1}^{N/2} e^{(\pi/4) \chi^{2j+1}_L \chi^{2j}_R} \ket{0},
\end{equation}
since $e^{-(\pi/4) \chi^{2j+1}_L \chi^{2j}_R} (\chi_L^{2j+1}) e^{(\pi/4) \chi^{2j+1}_L \chi^{2j}_R} = \chi_R^{2j}$ and $e^{-(\pi/4) \chi^{2j+1}_L \chi^{2j}_R} (\chi_R^{2j}) e^{(\pi/4) \chi^{2j+1}_L \chi^{2j}_R} = -\chi_L^{2j+1}$.

Both the EPR and TFD states can also be expressed in the energy eigenbases of the individual SYK models, $H_L$ and $H_R$.
To do so, we must first address two subtleties: defining complex conjugation for fermionic systems, and taking into account the fermionic anti-commutation between the left and right subsystem.
The former is handled by the anti-unitary  particle-hole transformation $P$~\cite{fidkowski2011topological,you2017sachdev,Cotler_black_2017}, defined such that 
\begin{equation}
    P^{-1} i P = -  i, \,\,\,\, P^{-1} \chi^j P =  \chi^j,
\end{equation}
with $P^2 = \pm 1$ dependent on the total number of fermions $N$.
The fermionic anti-commutation can be incorporated by defining the modified right operators, $\tilde{\chi}_R^j = F_L \chi_R^j$.
Here, $F_L = i^{N/2} \prod_{j=1}^N \chi^j_L$ is the fermion parity operator on the left subsystem, with eigenvalues $\pm 1$.
The modified right operators commute with all operators on the left subsystem.
This allows us to properly define a tensor product between the left subsystem, acted on by $\chi_L^j$, and a right subsystem, acted on by $\tilde{\chi}_R^j$.

With these definitions in hand, we can write the EPR state as follows.
Given any complete basis, $\{ \ket{v} \}$, for the individual left and right subsystems, we have
\begin{equation} \label{eq: EPR basis}
    \ket{\text{EPR}} = \frac{1}{2^{N/4}} \sum_v \ket{v} \otimes \ket{v^*},
\end{equation}
where we define,
\begin{equation} \label{eq: fermion cc}
    \ket{v^*} \equiv \Theta \ket{v}, \,\,\,\, \text{ with } \,\,\,\,  \Theta = P e^{i  (\pi/4) F}.
\end{equation}
The rotation $e^{i  (\pi/4) F}$ serves to append a fermion parity operator to each right fermion $\tilde{\chi}_R^j$; see below.
To show that the EPR state obeys Eq.~(\ref{eq: EPR condition}), let us consider the two states appearing in Eq.~(\ref{eq: EPR condition}).
The first state can be written,
\begin{equation}
    \chi_L^j \ket{\text{EPR}} 
    = \frac{1}{2^{N/4}} \sum_v ( \chi_L^j \ket{v} ) \otimes \ket{v^*} 
    = \frac{1}{2^{N/4}} \sum_{v,w} \bra{w} \chi^j \ket{v} \left(  \ket{w} \otimes \ket{v^*} \right).
\end{equation}
The second state can be written,
\begin{equation} \label{eq: chiRj EPR}
    \chi_R^j \ket{\text{EPR}} 
    = \frac{1}{2^{N/4}} \sum_w ( F_L \ket{w} ) \otimes ( \tilde{\chi}_R^j \ket{w^*} )
    = \frac{1}{2^{N/4}} \sum_{w,v} (-1)^{|w|} \bra{v^*} \chi^j \ket{w^*} \left(  \ket{w} \otimes \ket{v^*} \right),
\end{equation}
where $|w|$ denotes the number of fermions in state $\ket{w}$.
Since $\Theta$ is anti-unitary, we have
\begin{equation}
\begin{split}
    \bra{v^*} \chi^j \ket{w^*} 
    & = \bra{w} \Theta^{-1} \chi^j \Theta \ket{v} \\
    & = \bra{w} e^{-i(\pi/4)F} P^{-1} \chi^j P e^{i(\pi/4)F} \ket{v} \\
    & = \bra{w} e^{-i(\pi/4)F} \chi^j e^{i(\pi/4)F} \ket{v} \\
    & = \bra{w} -i F \chi^j \ket{v} \\ 
    & = -i (-1)^{|w|} \bra{w} \chi^j \ket{v}. \\ 
\end{split}
\end{equation}
Plugging this into Eq.~(\ref{eq: chiRj EPR}), we have $\chi_R^j \ket{\text{EPR}} = -i \chi_L^j \ket{\text{EPR}}$, which yields Eq.~(\ref{eq: EPR condition}).
In the fourth line above, we use that $e^{i (\pi/4) F} = (1+iF)/\sqrt{2}$ since $F^2 = 1$.

From the expression, Eq.~(\ref{eq: EPR basis}), for the EPR state, we can write the TFD state as
\begin{equation}
    \ket{\text{TFD}} = \frac{1}{\sqrt{\mathcal{Z}}} \sum_m e^{-\beta E_m / 2} \ket{m} \otimes \ket{m^*},
\end{equation}
where $\{ \ket{m} \}$ is the energy eigenbasis of the individual SYK model, with energies $E_m$.
We note that both $P$ and $F$ commute with the SYK Hamiltonian, which implies that $\ket{m^*}$ is an eigenstate with the same energy as $\ket{m}$.

\subsection{Large-$N$ equations of motion} \label{sec: SD eqs}

Similar to the original SYK model, the coupled Hamiltonian, Eq.~\eqref{eq:H_MQ}, is exactly solvable in the large-$N$ limit. 
At equilibrium, the solution is a system of closed equations for the two-point functions, known as the Schwinger-Dyson equations.
For quenched dynamics, the analogous solution is known as the Kadanoff-Baym equations. 
The derivations of these equations are described in detail in prior works \cite{maldacena2016remarks,Eberlein_quantum_2017,maldacena2021syk}; therefore, we provide only a brief summary of the results relevant to our study.

\emph{Equilibrium dynamics}---We begin by defining the two-point function under imaginary (Euclidean) time evolution,
\begin{equation}
    G_{A B}(\tau) \equiv \left <\mathcal{T} \chi_A^j(\tau) \chi^j_B(0) \right >_\beta
\end{equation}
where $A, B$ denotes the subsystem ($L$ or $R$),  $\mathcal{T}$ corresponds to imaginary-time ordering, and $\left < \cdots \right >_\beta = \textrm{Tr}[\cdots e^{-\beta H_{MQ}}]/\textrm{Tr}[e^{-\beta H_{MQ}}]$ is the thermal average. 
In the large-$N$ limit, one can solve for $G_{AB}(\tau)$ exactly by resumming a series of ladder diagrams or taking the saddle point of the large-$N$ path integral. 
Introducing the matrix form
\begin{equation}
  \textbf{G}(\tau) =  \begin{pmatrix}
G_{LL}(\tau) & G_{LR}(\tau) \\
G_{RL}(\tau) & G_{LL}(\tau)
\end{pmatrix} 
\end{equation}
and similarly for the other variables, the solution is \cite{maldacena2021syk}
\begin{equation} \label{eq: SD}
\begin{split}
\textbf{G} (\tau) &=  ( \bm{\Sigma} (\tau) - i \bm{\mu}(\tau)) * \bm{G} (\tau) + \bm{\delta}(\tau) \\
\bm{\Sigma} (\tau) &= J^2 \textbf{G}^3(\tau) 
\end{split}
\end{equation}
where 
\begin{equation}
\bm {\mu} (\tau) = \begin{pmatrix}
0 & \mu \delta(\tau) \\
-\mu \delta(\tau) & 0
\end{pmatrix} , \quad \bm {\delta} (\tau) = \begin{pmatrix}
\delta(\tau) & 0  \\
0 & \delta(\tau)
\end{pmatrix} .
\end{equation} 
Here, $*$ denotes convolution along the Euclidean circle ($0 \le \tau \le \beta$), and the cube in $\textbf{G}^3(\tau)$ is applied element-wise.
Note that taking $\mu = 0$ leads to independent equations for $G_{LL}(\tau)$ and $G_{RR}(\tau)$, corresponding to the Schwinger-Dyson equations for the uncoupled SYK model \cite{maldacena2016remarks}.

As described in Ref.~\cite{plugge2020revival}, it is convenient to decouple the above equations (with respect to the subsystem indices) by introducing the rotated basis,
\begin{equation}
\begin{split}
G_{\pm}(\tau) &=  G_{LL} (\tau) \pm i G_{LR}(\tau) \\
\Sigma_{\pm}(\tau) &=  \Sigma_{LL} (\tau) \pm i \Sigma_{LR}(\tau).
\end{split}
\end{equation}
Moreover, from the symmetries, $G_{LL}(\tau) = G_{RR}(\tau) $ and $G_{LR}(\tau) = -G_{RL}(\tau)$, one finds $G_\pm(\tau) = -G_\mp(-\tau)$.
As a result, Eq.~\eqref{eq: SD} can be simplified to a single pair of equations \cite{plugge2020revival}
\begin{equation} \label{eq: SD-imag}
\begin{split}
G_+(i\omega_n) &= \frac{1}{i \omega_n - \mu - \Sigma_+(i\omega_n)} \\ 
\Sigma_+(\tau) &= -\frac {J^2} 4 \left[3 G_+^2 G_+(-\tau) + G_+(-\tau)\right],
\end{split}
\end{equation}
where the first equation is now written in the frequency domain,
\begin{equation}
G_+(i\omega_n) = \int_0^\beta d\tau e^{i\omega_n \tau} G_+(\tau),
\end{equation}
and $\omega_n = (2n+1)\pi / \beta, n \in \mathbb{Z}$ is a Matusbara frequency.

By analytic continuation, one can derive analogous equations for the equilibrium two-point functions under real-time evolution \cite{maldacena2018eternal,plugge2020revival}. 
In particular, we introduce the Wightman and retarded two-point functions, respectively, as
\begin{equation}
\begin{split}
    G^>_{AB}(t)&\equiv -i\left < \chi_A^i(t) \chi^i_B(0) \right >_\beta \\
    G^R_{AB}(t) &\equiv \Theta(t) \left (G_{AB}^>(t) + G_{BA}^{>}(t)\right),
\end{split}
\end{equation}
where $\Theta(t)$ is the Heaviside function.
Under the fluctuation-dissipation theorem, these are related by
\begin{equation}
G_{LL}^> (\omega) = \frac{2 i \textrm{Im} G_{LL}^R(\omega)}{e^{-\beta \omega} + 1} , \quad G_{LR}^> (\omega) = \frac{\textrm{Re} G_{LR}^R(\omega)}{e^{-\beta \omega} + 1},
\end{equation}
where
\begin{equation}
G^>(\omega) = \int_{-\infty}^{\infty} dt e^{i\omega t} G^>(t), \quad G^>(t) = \int_{-\infty}^{\infty} \frac{dt}{2\pi} e^{-i\omega t} G^>(\omega).
\end{equation}
We also define the spectral functions,
\begin{equation}
\begin{split}
\rho_{LL}(\omega) &= \frac 1 2 [\rho_+(\omega)+\rho_+(-\omega)] \\
\rho_{LR}(\omega) &= \frac 1 2 [\rho_+(\omega)-\rho_+(-\omega)],
\end{split}
\end{equation}
and occupation functions,
\begin{equation}
n_{AB}(\omega) = \frac{\rho_{AB}(\omega)}{e^{-\beta \omega} + 1} .
\end{equation}
As above, the analysis is simplified by working in the rotated basis, for which $G_\pm^R(t) = G^R_{LL}(t) \pm G^R_{LR}(t)$, $G_\pm^R(t) = -G^R_{\mp}(-t)$, and $\rho_+(\omega) = -\frac 1 \pi \textrm{Im} G^R_+(\omega)$.
In this basis, the Schwinger-Dyson equations in real-time are given by \cite{plugge2020revival}
\begin{equation} \label{eq: SD-real}
\begin{split}
G_+^R(t) &= \frac{1}{\omega + i\epsilon - \mu -\Sigma_+^R(\omega)} \\
\Sigma_+^R(t) &= -2i J^2 (\textrm{Re} [n^3_{LL}(t)] - i \textrm{Im}[n^3_{LR}(t)]),
\end{split}
\end{equation}
where the first line utilizes the $i\epsilon$ prescription. 

In the following section, we describe an efficient iterative procedure for numerically solving both the real- and imaginary-time Schwinger-Dyson equations. 
The resulting solutions yield equilibrium properties for the coupled model and provide the initial conditions for simulating quenched dynamics.  

%

\emph{Quenched dynamics}---We next turn to the equations of motion for real-time, quenched dynamics, in which we treat $\mu$ as a time-dependent variable. 
Unlike the case of equilibrium, one cannot assume time-translation invariance and must consider two-point functions with respect to the time of each operator individually:
\begin{equation}
\begin{split}
G^>_{AB}(t_1, t_2) &\equiv -i\left < \chi_A^j(t_1) \chi^j_B(t_2) \right > \\
G^<_{AB}(t_1, t_2) & \equiv -i \left < \chi_A^j(t_2) \chi^j_B(t_1) \right > = -G_{BA}^>(t_2,t_1)  \\
G^R_{AB}(t_1,t_2) &\equiv \Theta(t_1-t_2) \left (G_{AB}^>(t_1,t_2) - G_{AB}^{<}(t_1,t_2)\right) \\
G^A_{AB}(t_1,t_2) &\equiv \Theta(t_2-t_1) \left (G_{AB}^<(t_1,t_2) - G_{AB}^{>}(t_1,t_2)\right).
\end{split}
\end{equation}
Using Keldyish formalism, one can derive the following system of equations \cite{maldacena2021syk}:
\begin{equation} \label{eq: KB}
\begin{split}
\partial_{t_1} \textbf{G}^> (t_1, t_2) = \bm{\mu}(t_1) \textbf{G}^> (t_1, t_2) - i \int_{-\infty}^{\infty} dt \left(\bm{\Sigma}^R(t_1, t) \textbf{G}^> (t, t_2) + \bm{\Sigma}^> (t_1, t) \textbf{G}^A(t, t_2)\right) \\
\partial_{t_2} \textbf{G}^> (t_1, t_2) = \bm{\mu}(t_2) \textbf{G}^> (t_1, t_2) - i \int_{-\infty}^{\infty} dt \left(\textbf{G}^R(t_1, t) \bm{\Sigma}^> (t, t_2) + \textbf{G}^> (t_1, t) \bm{\Sigma}^A(t_1, t)\right)
\end{split}
\end{equation}
where
\begin{equation}
\begin{split}
\Sigma_{AB}^>(t_1,t_2) &= -J^2 (G_{AB}^>(t_1,t_2))^3 \\
\Sigma_{AB}^<(t_1,t_2) &= -J^2 (G_{AB}^<(t_1,t_2))^3 \\
\Sigma_{AB}^R(t_1,t_2) &\equiv \Theta(t_1-t_2)(\Sigma_{AB}^>(t_1,t_2)-\Sigma_{AB}^<(t_1,t_2))\\
\Sigma_{AB}^A(t_1,t_2) &\equiv - \Theta(t_2-t_1)(\Sigma_{AB}^>(t_1,t_2)-\Sigma_{AB}^<(t_1,t_2)),
\end{split}
\end{equation}
and
\begin{equation}
\bm {\mu} (t) = \begin{pmatrix}
0 & \mu(t) \\
-\mu(t) & 0
\end{pmatrix} 
\end{equation}
is time-dependent. 
Note that, due to causality, the integrals in the above equations depend only on the solution for $\textrm{G}^>(t_1,t_2)$ at past times, i.e.~$t_1, t_2 < \textrm{max}(t_1,t_2)$.
Thus, given the solution for $\textrm{G}^>(t_1,t_2)$ in the range $t_1, t_2 < t$, one can solve for the derivatives at the boundary, $t_1 = t$ or $t_2 = t$, and propagate the solution forward in time. 
This property is essential for the numerical procedure described in the following section. 
%

%

\section{Details on numerical solutions to the large-$N$ equations of motion}

%

\subsection{Equilibrium dynamics}
To numerically solve the Schwinger-Dyson equations [Eq.~\eqref{eq: SD-imag} and ~\eqref{eq: SD-real}], we implement the iterative method introduced in Ref.~\cite{maldacena2016remarks}. 
Specifically, we begin with the free-fermion solution for the two-point function, given in imaginary- and real-time, respectively, as \cite{plugge2020revival}
\begin{equation}
G^0_+(i\omega_n) = \frac{1}{i\omega_n-\mu},  
\end{equation}
and
\begin{equation}
\quad G^0_+(\omega) = \frac{1}{\omega+i\epsilon-\mu},
\end{equation}
where $\epsilon$ is a small numerical value, e.g.~$10^{-5}$.
By performing discrete Fourier transforms between the time and frequency domain, we evaluate the Schwinger-Dyson equations and update the two-point function using a weighted sum of the initial and newly computed solutions. 
The steps are repeated until the two-point function converges to a desired precision.  
As the most computationally intensive step is a discrete Fourier transform, this algorithm can be performed efficiently with up to $\sim 10^7-10^8$ points of time discretization. 
This ensures very high accuracy for all equilibrium results presented in this work.


Based on the above approach, we solve for both the real-time and imaginary-time dynamics of the coupled model, as well as for the uncoupled SYK model. 
From the solutions, we compute the following quantities:
\begin{itemize}
\item The two-sided correlator, $G_{LR}$, for the TFD state at inverse temperature $\beta$. 
This is equivalent to evaluating the imaginary-time correlator of the uncoupled SYK model (at inverse temperature $\beta$), i.e.~$G_{LR} \equiv G_\textrm{SYK}(\beta/2)$.
\item The two-sided correlator, $G_{LR}$, for the ground-state of the coupled model. 
Although the SD equations can only be solved numerically at finite temperatures, we can approximate the ground state by evaluating at a low temperature compared to the energy scales $\mu$ and $J$; specifically, for the ground-state results shown in Fig.~\ref{fig: ground state}(a), we use $\beta = 30/\mu$. 
The two-sided correlator is then simply $G_{LR}(0)$.
\item The expectation value for the single-side energy, $E \equiv \langle H_L \rangle$.
For the uncoupled SYK model, a convenient expression is \cite{maldacena2021syk}
\begin{equation}
\frac E N = - \frac{J^2}{4} \int_0^\beta d\tau \; G_{\textrm{SYK}}^4(\tau).
\end{equation}
Similarly, for the coupled model, we have
\begin{equation} \label{eq: E_L}
\frac E N = - \frac{J^2}{2} \int_0^\beta d\tau \; (G_{LL}^4(\tau) + G_{LR}^4(\tau)).
\end{equation}
\item The gap of the coupled system, $\Delta_m$, with respect to ``matter'' excitations (i.e.~obtained by acting $\chi_L^j$ on the ground state). 
Recall that the spectral decomposition of the two-point function is $G_{LL}(\tau) = \sum_n e^{-\tau (E_n - E_0)} |\bra{n}\chi^j_L \ket{\textrm{GS}}|^2$ for eigenstates $\ket{n}$ with energy $E_n$.
At late times, the decay is dominated by the minimum energy gap, $G_{LL}(\tau) \sim e^{-\tau(E_1-E_0)}$.
Thus, by fitting the late-time decay to a simple exponential, we can determine $\Delta_m = E_1 - E_0$.
\item The ``graviton'' gap, $\Delta_g$, based on the eigenstate thermalization framework. As described in Eq.~\eqref{eq: sho freq}, the gap can be predicted from equilibrium properties of the \emph{uncoupled} SYK model, namely $G_\beta \equiv G_\textrm{SYK}(\beta/2)$ and $G^{\prime \prime}_\beta \equiv \left . \partial ^2 G_\textrm{SYK}(\beta/2-it)/\partial t^2 \right |_{t=0}$. 
For the latter, we compute $G_\textrm{SYK}$ at complex times by performing a numerical analytic continuation of $G^>_\textrm{SYK}(t)$.
\end{itemize}

\subsection{Quenched dynamics}
Following the approach of Ref.~\cite{Eberlein_quantum_2017} and \cite{maldacena2021syk}, we numerically solve the Kadanoff-Baym equations by propagating the two-point function forward in time using a predictor-corrector scheme \footnote{We sincerely thank Alexey Milekhin for sharing his code with us and for helpful discussions regarding the propagation method.}. 
In all cases, we consider the initial state of the the system, at $t = 0$, to be the equilibrium state of the coupled system at a specified inverse temperature $\beta$ and coupling strength $\mu$.
This is implemented by initializing the two-point function as the solution of the Schwinger-Dyson equations, $\textbf{G}^>(t_1,t_2) = \textbf{G}^>(t_1-t_2)$, over an initial time window, $-t_\textrm{init} \le t_1, t_2 \le 0$.

Based on these initial conditions and a specified time-dependent coupling $\mu(t)$, we propagate the solution forward in time by intervals of $\Delta t$ using the following procedure. 
Starting with the $t_1$ direction, we first define the right-hand side of the first equation in Eq.~\eqref{eq: KB} as
\begin{equation}
F_1(\textbf{G}^>; t_1, t_2) = \bm{\mu}(t_2) \textbf{G}^> (t_1, t_2) - i \int_{-\infty}^{\infty} dt \left(\textbf{G}^R(t_1, t) \bm{\Sigma}^> (t, t_2) + \textbf{G}^> (t_1, t) \bm{\Sigma}^A(t, t_2)\right),
\end{equation}
and calculate an initial prediction for the update as
\begin{equation}
\textbf{G}^{>,(p)}(t_1+\Delta t, t_2) =  \textbf{G}^>(t_1, t_2) + \Delta t \; F_1(\textbf{G}^>; t_1, t_2)),
\end{equation}
where the integral in $F_1$ is performed using the trapezoidal rule. 
Next, we correct the update using
\begin{equation}
\textbf{G}^{>}(t_1+\Delta t, t_2) =  \textbf{G}^>(t_1, t_2) + \frac{\Delta t}{2} \; \left[ F_1(\textbf{G}^>; t_1, t_2) + F_1(\textbf{G}^{>,(p)}; t_1 + \Delta t, t_2) \right].
\end{equation}
While the correction can be improved with further iteration, we generally find that a single step is sufficient.
Updating along the $t_2$ direction proceeds analogously, where we define the second equation in Eq.~\eqref{eq: KB} as $F_2(\textbf{G}^>; t_1, t_2)$. 
Finally, for the ``diagonal'' update, we calculate the initial prediction as
\begin{equation}
\textbf{G}^{>,(p)}(t_1+\Delta t, t_2+\Delta t) =  \textbf{G}^>(t_1, t_2) + \Delta t \; \left [F_1(\textbf{G}^>; t_1 + \Delta t, t_2) + F_2(\textbf{G}^>; t_1, t_2+ \Delta t)\right]
\end{equation}
and update the prediction with
\begin{equation}
\begin{split}
\textbf{G}^{>}(t_1+\Delta t, t_2+\Delta t) =  \textbf{G}^>(t_1, t_2) + \frac{\Delta t}{2} \; \left [F_1(\textbf{G}^>; t_1, t_2) + F_2(\textbf{G}^>; t_1, t_2) \right. \\
\left .+F_1(\textbf{G}^{>,(p)}; t_1+\Delta t, t_2+\Delta t) + F_2(\textbf{G}^{>,(p)}; t_1 + \Delta t, t_2+\Delta t)\right].
\end{split}
\end{equation}
By incrementing in this fashion, we solve for $G^>(t_1,t_2)$ over a total time range, $0 < t_1, t_2 < T$. 

From the numerical solutions, we immediately determine the single-sided and two-sided correlators,  $G_{LL}$ and $G_{LR}$, as a function of time.
Analogous to Eq.~\eqref{eq: E_L}, we can also compute the energy of the single-sided energy, at time $t$, with \cite{maldacena2021syk}
\begin{equation}
\frac {E} N = - i\frac{J^2}{4} \int_{-\infty}^{\infty} dx \; \left(G_{t,{LL}}^4(x)+G_{t,{LR}}^4(x)\right), 
\end{equation}
where 
\begin{equation}
G_{t,{AB}}(x) = \Theta(x) G^>_{AB}(t-x,t) + \Theta(-x) G^>_{AB}(t,t+x).
\end{equation}
%

Compared to solving the Schwinger–Dyson equations, the above algorithm has a significantly higher computational cost.
For a time discretization with $M$ grid points along each time direction, the time cost of each individual update scales as $\sim M$, leading to an overall cost of $\sim M^3$ for the full algorithm---much larger than the $\sim M \log M $ scaling in the case of the equilibrium calculations. 
In our implementation, we perform the convolutions in Eq.~\eqref{eq: KB} via matrix multiplication on a GPU; this approach enables simulations with up to $M \sim 8000$ time steps in approximately $20 $ hours.

Due to the limited number of time steps, we must be more careful to choose simulation parameters to minimize numerical errors.
Specifically, two conditions should be satisfied: 
\begin{itemize}
\item The step size $\Delta t$ must be small compared to the high-frequency behavior of the system. 
\item The initial time window, defined by $t_\textrm{init}$, must exceed the decay time of the equilibrium two-point functions, $\textbf{G}^>(t)$; otherwise, the integrals in Eq.~\eqref{eq: KB} become inaccurate. 
\end{itemize}
Crucially, at low temperatures, the two-point functions of the coupled system exhibit long-lived oscillations---corresponding to wormhole revivals in the small-$\mu$ limit \cite{maldacena2003eternal,plugge2020revival}. 
Therefore, the latter condition imposes a limitation on the range of initial temperatures that can be accurately simulated.
Similarly, we cannot initialize the simulations at infinite $\mu$, as the two-point functions would be perfectly sinusoidal.

\begin{figure*}[t]
\centering
\includegraphics[width=\linewidth]{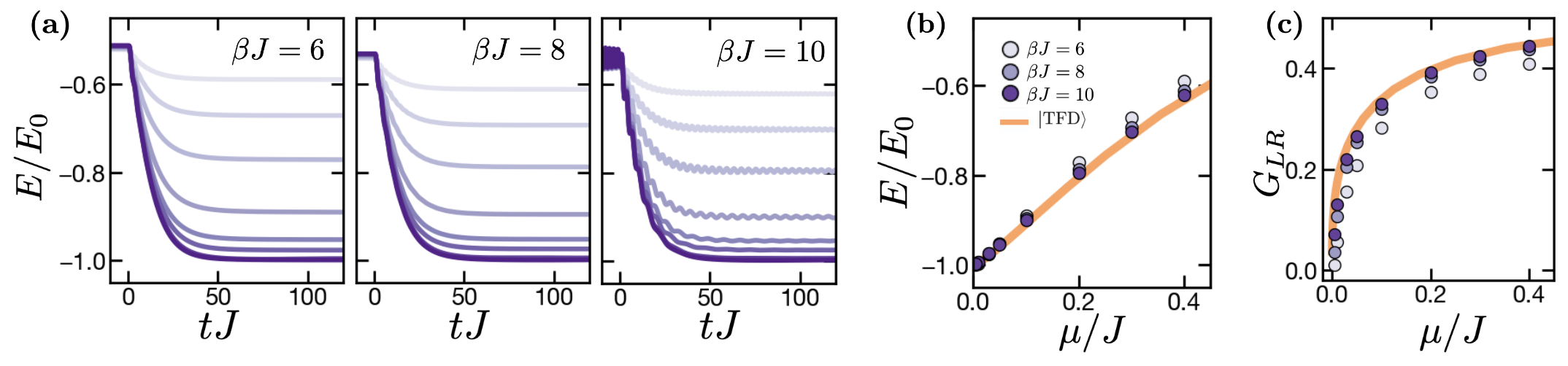}
\caption{Comparison of the adiabatic cooling protocol shown in Fig.~\ref{fig: adiabatic}(b) of the main text under three different initial temperatures, $\beta = 6,8,$ and $10$. (\textbf{a}) The single-side energy $E = \langle H_L \rangle $ as a function of time, for a coupling that decreases adiabatically as $\mu(t) = (\mu_i-\mu_f) e^{-t/T_*}+\mu_f$, with $T_* = 10/J$ and various initial couplings $\mu_i$. (\textbf{b}) As the initial temperature decreases, the final single-side energy, evaluated at $t = 225/J$, approaches the expected value of the TFD state. (\textbf{c}) Similar convergence is seen in the final value of the two-sided coupling, $G_{LR}$. The simulations use a step size $\Delta t = 0.05/J$ and an initial time range $t_\textrm{init}=80/J$.}
\label{fig: temperature}
\end{figure*}

In what follows, we present additional results for the quenched protocols shown in the main text. 
In Fig.~\ref{fig: temperature}, we analyze the effect of the initial temperature on the adiabatic cooling protocol.
In particular, we consider the same protocol as shown in Fig.~\ref{fig: adiabatic}(b) under three different initial temperatures, $\beta = 6,8,$ and $10$.
As the initial temperature decreases, the final values for the single-sided energy and two-sided correlation show small but systematic improvements relative to the expected values from the TFD state [Fig.~\ref{fig: temperature}(b-c)].
This indicates that finite-temperature corrections are minimal and that our simulations approximately capture the behavior of the protocol initialized in the ground state. 
We note that, for the lowest temperature, small oscillations are present in the time-profile of the single-sided energy, including at early and late times at which the coupling $\mu$ is constant. 
We conclude that this temperature is outside of the range of validity for our simulations, namely, due to insufficient decay in the initial two-point functions.
To avoid such errors yet approximate the ground-state behavior, the results shown in Fig.~\ref{fig: adiabatic}(b-d) of the main text were obtained using an initial temperature of $\beta = 8$.   

\begin{figure*}[t]
\centering
\includegraphics[width=\linewidth]{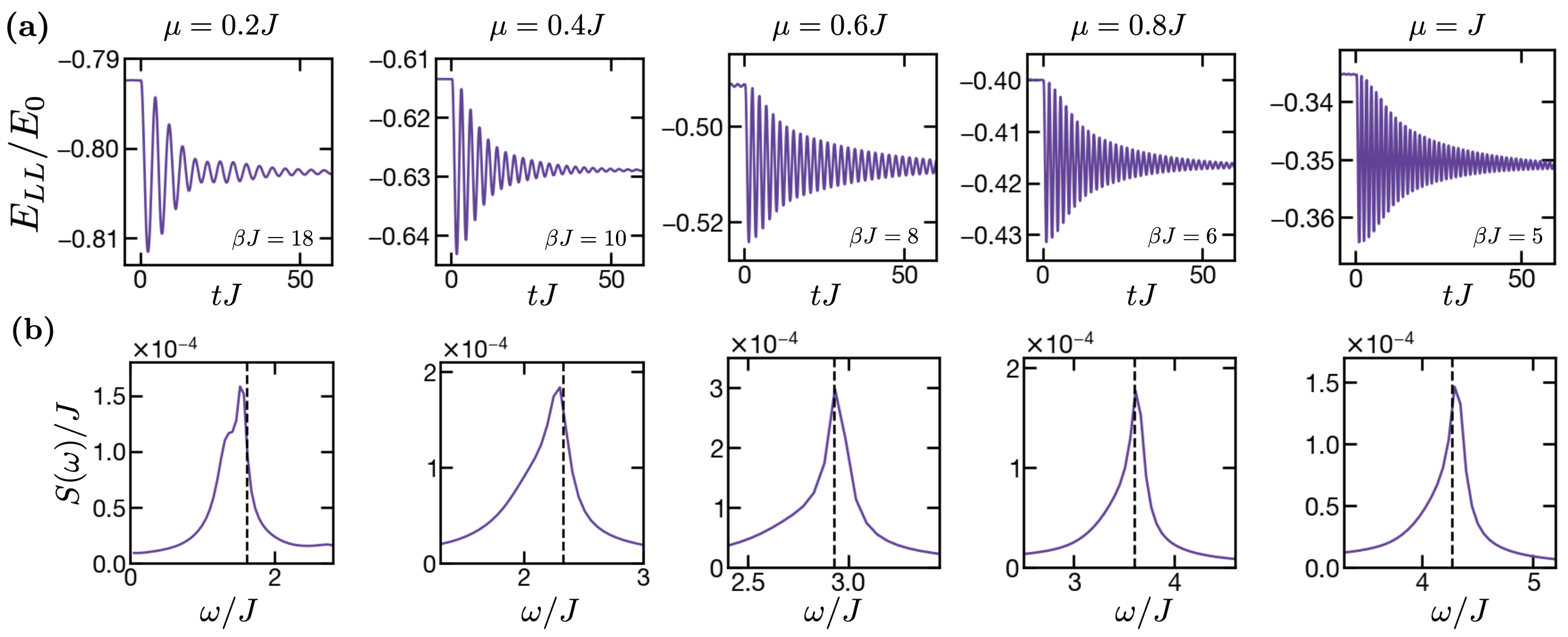}
\caption{Analysis of the ``graviton'' gap, $\Delta_g$, using the protocol shown in Fig.~\ref{fig: adiabatic}(a) of the main text. The system is initialized in a low-temperature state at coupling strength  $\mu_i = \mu$. At $t=0$, the coupling strength is abruptly reduced to $\mu_f =  0.95 \mu$. (\textbf{a}) This quench leads to persistent oscillations in the time-profile of the single-side energy, $E \equiv \left <H_L\right>$. (\textbf{b}) To determine the characteristic oscillation frequency, we compute the spectrum, $S(\omega) =  \int_0^T e^{-i\omega t} \left |E(t) \right |^2 dt$ for $TJ = 120$. For each initial coupling $\mu$, the spectrum exhibits a sharp resonance, whose peak frequency is interpreted as the ``graviton'' gap, $\Delta_g$. In all cases, the peak frequency agrees well with the predicted $\Delta_g$ from the eigenstate thermalization framework, given by Eq.~\eqref{eq: sho freq} and evaluated at $\mu_f$ (vertical dashed lines). In these simulations, the step size is $\Delta t = 0.05/J$, the initial time range is $t_\textrm{init} = 80/J$, and, for each coupling strength $\mu$, the initial temperature $\beta$ is chosen to be the lowest temperature without observable numerical errors.}
\label{fig: oscillations}
\end{figure*}

In Fig.~\ref{fig: oscillations}, we present our analysis for the ``graviton'' gap, $\Delta_g$, shown in Fig.~\ref{fig: adiabatic}(a) of the main text. 
The results are based on the ``sudden'' quench protocol, in which the coupling is reduced at $t=0$ from $\mu_i = \mu$ to $\mu_f = 0.95 \mu$.
As shown in Fig.~\ref{fig: oscillations}(a), this change leads to oscillations in the single-side energy, $E$. 
To characterize the dominant frequency, we take a Fourier transform of $E(t)$ for $t > 0$ and determine the peak frequency of the resonance in the resulting spectrum [Fig.~\ref{fig: oscillations}(b)]. 
For all coupling strengths, this frequency shows close agreement with the predicted gap from Eq.~\eqref{eq: sho freq}. 

\section{Theoretical framework based on eigenstate thermalization} \label{app: ETH}

In this section, we provide additional details on our heuristic theoretical framework to bridge the large and small $\mu$ limits of the coupled Hamiltonian.
As mentioned in the main text, our framework is based on an ansatz, that the essential physics of the coupled Hamiltonian can be captured through its action on a restricted subspace of energy eigenstates of the uncoupled SYK models.
The action of the coupled Hamiltonian on this subspace can then be analyzed using the eigenstate thermalization hypothesis (ETH).
This framework was introduced in Ref.~\cite{cottrell2019build} and analyzed for conformal field theories.
In what follows, we begin in Section~\ref{sec: ETH review} by providing a short summary of the ETH.
We then turn to the coupled Hamiltonian, and provide a succinct derivation of the effective Schrodinger equation of Ref.~\cite{cottrell2019build} (Sections~\ref{sec: ETH discrete} and~\ref{sec: ETH continuous}).
In Section~\ref{sec: SYK ETH}, we derive the parameters of the Schrodinger equation in the specific case of the SYK model, and verify that it agrees exactly with the dynamics of the ``graviton'' mode found in Ref.~\cite{maldacena2018eternal}.
Finally, in Section~\ref{sec: accuracy}, we provide a simple self-consistency check for the accuracy of the theoretical framework, which suggests that the framework accurately captures ground state properties of the coupled SYK Hamiltonian in both the large and small $\mu$ limits.

\subsection{Review of the eigenstate thermalization hypothesis} \label{sec: ETH review}

Consider a many-body Hamiltonian $H$ with eigenstates $\ket{m}$ and energies $E_m$.
The ETH posits that the matrix elements of few-body operators in the energy eigenbasis are described as follows.
First, the diagonal matrix elements are given by the expectation value of the operator in the corresponding thermal state,
\begin{equation}
    \bra{m} A \ket{m} \approx \tr( \rho_{\beta_m} A ).
\end{equation}
where $\rho_{\beta_m} = e^{-\beta_m H}/\tr(e^{-\beta_m H})$ is the thermal density matrix at inverse temperature $\beta_m$ such that $\tr( \rho_{\beta_m} H ) = E_m$.
When $A$ is a fermionic operator, the diagonal elements are zero since we can choose $\ket{m}$ to have fixed fermion parity.
Second, the off-diagonal matrix elements are given by
\begin{equation}
\bra{m} A \ket{n} = e^{-S_{mn}/2} \,  f_{mn} \,  \R^A_{mn}.
\end{equation}
Here, $S_{mn} = S(E_{mn})$ is the entropy of a thermal state at inverse temperature $\beta_{mn}$, where $\beta_{mn}$ is set by the average energy, $E_{mn} = (E_m+E_n)/2$.
The function $f_{mn} = f(E_{mn},\omega_{mn})$ is a smooth function of the average energy, $E_{mn}$, and the energy difference, $\omega_{mn} = E_m - E_n$.
The $\R^A_{mn}$ are random variables with zero mean and unit variance\footnote{For completeness, we note that higher moments of these random variables are not necessarily independent between different $i, j$~\cite{jafferis2023matrix,jafferis2209jt,belin2023approximate}. This is not important for the current analysis, which will only involve the  moment, $| \R^A_{mn} |^2$.}.

The function $f(E_{mn},\omega_{mn})$ is set by the auto-correlation function of the operator $A$.
Decomposing $A$ in the energy eigenbasis, we have
\begin{equation}
G_\beta(t) = \text{tr}( A(t)  \rho^{1/2} A(0) \rho^{1/2}) = \frac{1}{\calZ} \sum_{mn} e^{-\beta E_{mn}} e^{i \omega_{mn} t} | A_{mn} |^2 \approx \frac{1}{\calZ} \sum_{mn} e^{-\beta E_{mn}} e^{i \omega_{mn} t} e^{-S(E_{mn})} |f_{mn}|^2  \\ 
\end{equation}
Here, we work with a particular thermal regularization of the auto-correlation function for later convenience (The analogous expression for the usual auto-correlation function can be obtained by shifting $t \rightarrow t + i \beta/2$.).
In the final step, we assume that the squared random variables can be approximated by their average, i.e.~$| \R^A_{mn}|^2 \approx 1$, since the sum is over exponentially many such variables. 

Changing coordinates from $E_m, E_n$ to $E_{mn}$, $\omega_{mn}$, and converting the sums to integrals, we have:
\begin{equation}
\begin{split}
G_\beta(t) & = \frac{1}{\calZ} \int dE e^{-\beta E + S(E)} \int d\omega  \left[ e^{S(E+\omega/2)+S(E-\omega/2)-2S(E)} \right] e^{i\o t} | f(E,\o)|^2 \\ 
& \approx \int d\omega  \left[ e^{S(E_\beta+\omega/2)+S(E_\beta-\omega/2)-2S(E_\beta)} \right] e^{i\o t} | f(E_\beta,\o)|^2 \\ 
 & \approx \int d\omega  e^{i\o t} | f(E_\beta,\o)|^2. \\ 
\end{split}
\end{equation}
In the second line, we use the standard saddle point approximation for the thermodynamic limit.
In the third line, we note that the difference of entropies vanishes to leading order in $\omega^2/N$.
From the final expression, we see that the function $|f(E_{mn},\omega_{mn})|^2$ is simply the Fourier transform of the auto-correlation function.
For local Hamiltonians, the support of the Fourier transform is tightly constrained within the energy window $| \omega | \lesssim \beta^{-1}$; outside of this window, the Fourier transform generally decays exponentially, $|f(E_\beta,\omega)| \sim \! e^{-\beta \omega}$.

\subsection{Matrix elements of the coupled Hamiltonian} \label{sec: ETH discrete}

Let us now turn to the coupled Hamiltonian,
\begin{equation}
H = H_L + H_R + \mu \sum_{j=1}^N i \chi^{j}_{L} \chi^{j}_{R},
\end{equation}
Our analysis in this subsection and the following subsection follows Ref.~\cite{cottrell2019build}, adapted to fermionic systems.

As introduced in the main text, we consider the basis of states, $\{ \ket{m n^*} \equiv \ket{m} \otimes \ket{n^*}\}$, where $\ket{m}$ and $\ket{n}$ are eigenstates of the SYK model and the complex conjugation, $\ket{n^*}$, is defined in Eq.~(\ref{eq: fermion cc}).
The matrix elements of $H_L$ and $H_R$ in this basis are diagonal in $m,n$,
\begin{equation}
    \bra{m' n'^*} H_L + H_R^* \ket{m n^*} = (E_m + E_n) \cdot \delta_{m m'} \delta_{n n'}.
\end{equation}
Meanwhile, the matrix elements of the coupling can be expressed in terms of the off-diagonal matrix elements of the local fermion operators,
\begin{equation}
    \bra{n' n^*} i \chi_L^j \chi_R^j \ket{m' m^*} = i \bra{n'} \chi^j F \ket{m'} \bra{n^*} \chi^j \ket{m^*}  = (-1)^{|m|+|m'|} \bra{n'} \chi^j \ket{m'} \bra{n} \chi^j \ket{m},
\end{equation}
following the analysis in Section~\ref{sec: fermion TFD}.
In general, these matrix elements will have wildly varying phases for different $m,n,m',n'$.
This drastically simplifies if we restrict attention to the ``paired'' subspace of eigenstates, with $m = n$ and $m' = n'$.
In this subspace, the matrix elements are entirely positive,
\begin{equation}
    \bra{n n^*} i \chi_L^j \chi_R^j \ket{m m^*} = \left|  \bra{n} \chi^j \ket{m} \right|^2,
\end{equation}
where we replace the index $m' \rightarrow n$ for convenience.
Invoking the ETH and taking the sum over all couplings $j$, we have
\begin{equation}
\begin{split}
\bra{nn^*} \mu  \sum_{j=1}^N i \chi_L^j \chi_R^j \ket{mm^*} = \mu  e^{-S_{mn}} \sum_{j=1}^N  | f^j_{mn} |^2  | \R^{j}_{mn} |^2 \approx \mu  e^{-S_{mn}} \sum_{j=1}^N  | f^j_{mn} |^2, \\
\end{split}
\end{equation}
where we approximate $| \R^{j}_{mn} |^2 \approx 1$ by its typical value.

Following the main text and Ref.~\cite{cottrell2019build}, we proceed by analyzing the action of the coupled Hamiltonian solely within the paired subspace.
In principle, one might worry that this restriction is not valid, since paired states,~$\ket{mm^*}$, can couple to unpaired states, ~$\ket{mn^*}$ with $m \neq n$, under the  coupled Hamiltonian.
Nevertheless, as discussed in the main text, we find that this restriction leads to remarkably accurate predictions for the coupled SYK Hamiltonian.
Along these lines, we refer to the subsequent Section~\ref{sec: SYK ETH} for a detailed comparison between the paired-subspace Hamiltonian and the graviton dynamics of the SYK model at small $\mu$, and Section~\ref{sec: accuracy} for our self-consistency calculation that supports the paired-subspace approximation at both small and large $\mu$.
%

In summary, within the paired subspace, the coupled Hamiltonian has matrix elements,
\begin{equation} \label{eq: matrix elements diag}
    H_{nn^*,mm^*} = 2 E_m \delta_{mn} + \mu N e^{-S_{mn}} | f_{mn} |^2,
\end{equation}
where we denote the average auto-correlation function as $| f_{mn} |^2 \equiv \frac{1}{N} \sum_{j=1}^{N} | f^j_{mn} |^2$.
The first term resembles an on-site potential energy, while the second term contains hoppings between nearby paired states.
In the next section, we will derive a continuum approximation for this Hamiltonian and show that it corresponds to a one-dimensional semi-classical harmonic oscillator.

\subsection{Continuum approximation for the coupled Hamiltonian} \label{sec: ETH continuous}

To take the continuum limit, we assume that the wavefunction of our state takes the form
\begin{equation}
\ket{\psi} = \sum_m e^{-S(E_m)/2} \cdot \psi(E_m) \ket{ mm^*},
\end{equation}
where the amplitudes $\psi(E_m)$ vary smoothly as a function of the energy $E_m$.
The prefactor $e^{-S(E_m)/2}$ corresponds to the inverse square root density of states, which guarantees that the continuum wavefunction, $\psi(E)$, is normalized, 
\begin{equation}
    \int dE \, |\psi(E)|^2 \approx \sum_m e^{-S(E_m)} \cdot |\psi(E_m)|^2 = \braket{\psi}{\psi} = 1.
\end{equation}

To derive the Schrodinger equation for $\psi(E)$, we take the time-derivative, $i\partial_t \ket{\psi} = H \ket{\psi}$, and restrict attention to the paired subspace, yielding
\begin{equation}
\begin{split}
i \sum_m e^{-S(E_m)/2} \cdot \partial_t \psi(E_m) \ket{ mm^*} & = \sum_{i} e^{-S(E_m)/2} \cdot \psi(E_m) \cdot \sum_n H_{nn^*,mm^*} \ket{ nn^*}. \\
\end{split}
\end{equation}
Inserting Eq.~(\ref{eq: matrix elements diag}) for the Hamiltonian matrix elements, the wavefunction evolves via
\begin{equation}
\begin{split}
i \partial_t \psi(E_m) & = 2E_m \cdot \psi(E_m) - \mu N \sum_{n} e^{\frac{1}{2} ( S_m - S_n - 2 S_{mn})} |f_{mn}|^2 \cdot \psi(E_n). \\
\end{split}
\end{equation}
Approximating the sum over $m$ as an integral, we have
\begin{equation}
\begin{split}
i \partial_t \psi(E_m) & \approx 2E_m \cdot \psi(E_m) - \mu N \int dE_n \, e^{\frac{1}{2} ( S_m + S_n - 2S_{mn})}  |f_{mn}|^2 \cdot \psi(E_n) \\
& \approx 2E_m \cdot \psi(E_m) - \mu N \int d\o_{mn} \left| f(E_m + \o_{mn}/2,\o_{mn}) \right|^2 \cdot \psi(E_m+\o_{mn}).  \\
\end{split}
\end{equation}
In the second line we switch integration coordinates to $\o_{mn} = E_m - E_n$, and use $e^{\frac{1}{2} ( S_m + S_n - 2S_{mn})} \approx e^{-\omega_{mn}^2 \beta^2 / 8 c N} \approx 1$. Here, $c = \mathcal{O}(1)$ is the specific heat, and the approximation follows because $\omega_{mn} \beta \lesssim \mathcal{O}(1)$ since $f(E_{mn}+\omega_{mn}/2,\omega_{mn})$ is exponentially suppressed at larger $\omega_{mn}$.
Finally,  we can Taylor expand $\psi(E_m+\omega_{mn}) \approx \psi(E_m) + \omega_{mn} \partial_{E_m}\psi(E_m) + (\omega_{mn}^2/2) \partial^2_{E_m} \psi(E_m)$ to obtain an effective Schrodinger equation,
\begin{equation}
\begin{split}
i \partial_t \psi(E) & \approx 2E \cdot \psi(E) - \mu N \left[ a(E) \psi(E) + b(E) \partial_{E} \psi(E) + c(E) \partial_E^2 \psi(E) \right],
\end{split}
\end{equation}
with coefficients,
\begin{equation}
\begin{split}
a(E) & = \int d\o  \left| f(E + \o/2,\o) \right|^2 \approx \int d\o  \left| f(E,\o) \right|^2 = G_{\beta(E)}(0), \\
b(E) & = \int d\o \, \o \cdot \left| f(E + \o/2,\o) \right|^2 \approx \int d\o \, \o^2 \cdot f(E,\o) \d_E f(E,\o) = \mathcal{O}\left( \frac{1}{\beta N} \right), \\
c(E) & = \frac{1}{2} \int d\o \, \o^2 \cdot \left| f(E + \o/2,\o) \right|^2 \approx  \frac{1}{2} G^{''}_{\beta(E)}(0). \\
\end{split}
\end{equation}
The linear term is suppressed by $1/N$ because $f(E,\o)$ is even in $\o$ (due to time-reversal symmetry), and each derivative applied to the first argument of $f$ gives a factor of $1/N$.
In the final line, we use double primes to denote the second time-derivative of $G^{''}_{\beta(E)}$, evaluated at $t=0$.

To write this equation in a nicer form, let us switch coordinates to the \emph{energy density}, $e \equiv E/N$, and drop the linear term in the thermodynamic limit.
This gives
\begin{equation}
\begin{split}
i N^{-1} \partial_t \psi(e) & \approx \left[ 2 e - \mu G_{\beta(e)} \right] \cdot \psi(e) -\mu G^{''}_{\beta(e)} \frac{ N^{-2} \d_e^2 }{2} \psi(e) .
\end{split}
\end{equation}
This is the usual single-particle Schrodinger equation for a particle in a potential $V(e) = 2e - \mu G_{\beta(e)}$, with a position-dependent mass, $m = ( \mu G^{''}_{\beta(e)} )^{-1}$.
The effective Planck's constant is $N^{-1}$, which is small in the thermodynamic limit.
We expect this Schrodinger equation to describe the low-energy physics in the paired subspace of states.

We can further simplify the effective Schrodinger equation by Taylor expanding the potential about its  minimum.
The minimum occurs at an energy density, $e^*$, determined by the solution to the equation,
\begin{equation} \label{eq: set e min}
    2 = \mu \partial_e G_{\beta(e)} \big|_{e=e^*},
\end{equation}
which is controlled by the value of $\mu$.
Nearby the minimum, we can expand the potential to quadratic order,
\begin{equation}
    2e - \mu G_{\beta(e)} \approx \text{constant} + \frac{\mu \partial_e^2 G_{\beta(e)} \big|_{e=e^*}}{2}  \, e^2,
\end{equation}
and we can also approximate the mass by its value at the potential minimum.
Thus, nearby the potential minimum, the behavior of the paired subspace is described by the simple harmonic oscillator,
\begin{equation} \label{eq: QSHO}
\begin{split}
i N^{-1} \partial_t \psi(e) & \approx \left( \text{constant} + \frac{ 1 }{2} k(\mu) (e-e^*)^2 \right) \psi(e) - \frac{ N^{-2} }{2 m(\mu)} \d_e^2 \psi(e).
\end{split}
\end{equation}
with spring constant $k(\mu) = \mu \partial^2_e G_{\beta(e)} \big|_{e=e^*}$ and mass $m(\mu) = ( \mu G^{''}_{\beta^*} )^{-1}$.
Here we denote the inverse temperature at the minimum as $\beta^* = \beta(e^*)$.

Applying standard formulas for the quantum harmonic oscillator, the ground state of the Hamiltonian at coupling strength $\mu$ is
\begin{equation} \label{eq: gs wavefunction}
    \psi_{\text{gs}}(e; \mu) = \left( \frac{N m(\mu) \omega(\mu)}{\pi} \right)^{1/4} \exp( - \frac{N m(\mu) \omega(\mu)}{2} (e - e^*)^2 ),
\end{equation}
where we define the frequency,
\begin{equation} \label{eq: sho freq}
    \omega(\mu) = \sqrt{\frac{k(\mu)}{m(\mu)}} = \mu \sqrt{\partial^2_e G_{\beta(e)} \big|_{e=e^*} \cdot G^{''}_{\beta^*}}.
\end{equation}
The ground state has an average energy density, $e^*(\mu)$, with respect to the individual SYK Hamiltonians, and a standard deviation, $\sqrt{N m(\mu) \omega(\mu)}$.
The Hamiltonian has a gap,
\begin{equation}
    \Delta_{g}(\mu) = N \left( N^{-1} \omega(\mu) \right) = \omega(\mu),
\end{equation}
where we have multiplied by $N$ to convert to our original energy units.
The excited states of the Hamiltonian are given by the excited states of the quantum harmonic oscillator\footnote{We note that this discussion only concerns excited states within the paired subspace. As previously mentioned, these correspond to so-called graviton excitations in the SYK model~\cite{maldacena2018eternal}. The effective Schrodinger equation does not capture matter excitations in the SYK model~\cite{maldacena2018eternal}, which have support on ``unpaired'' states.}.

Finally, we can quantify the closeness of the ground state to the desired TFD state.
The TFD state has a wavefunction
\begin{equation}
    \psi_{\text{tfd}}(e; \beta) = \frac{e^{(S(e) - \beta N e)/2}}{\tr(e^{-\beta H})} \approx \left( \frac{N \beta^2}{c} \right)^{1/4} \exp(- \frac{ N \beta^2 }{2c} (e-e_{\beta})^2),
\end{equation}
where we obtain the latter expression by Taylor expanding the entropy, where, again, $c$ is the specific heat at inverse temperature $\beta$.
The TFD state has a Gaussian wavefunction, with average energy density $e_\beta$ and standard deviation $\sqrt{N\beta^2/c}$.
For a desired inverse temperature $\beta$, we can maximize the overlap of the ground state with the TFD state by setting $\mu$ such that $e^*(\mu) = e_\beta$.
With this choice, the two states have an $\mathcal{O}(1)$ many-body overlap.
(In general, the overlap is not exactly one, because the widths of the two wavefunctions differ by an $\mathcal{O}(1)$ factor.)
We caution that, in reality, we expect corrections to the paired-subspace approximation to cause the many-body overlap  to instead decay exponentially in system size~\cite{maldacena2018eternal}.
Nonetheless, we expect local observables and correlation functions in the ground state to remain close to those in the TFD state, as has been verified at low temperatures in the SYK model~\cite{maldacena2018eternal}.
We refer to the main text and Section~\ref{sec: accuracy} for  related discussions.


\subsection{Parameters for the SYK model and comparison to known results} \label{sec: SYK ETH}

Let us now specialize our analysis to the SYK Hamiltonian.
In particular, this will allow us to benchmark our approximation by comparing the parameters of our effective Schrodinger equation with known SYK results at low temperatures, $\beta^{-1}$.
At low temperatures, the auto-correlation function of the SYK model is~\cite{maldacena2016remarks},
\begin{equation}
    G_\beta(t) = c_\Delta \left( \frac{\pi}{\mathcal{J} \beta \cosh( \pi t / \beta )} \right)^{2 \Delta}, \,\,\,\,\, c_\Delta = \frac{1}{2} \left[ (1-2\Delta) \frac{\tan(\pi \Delta)}{\pi \Delta} \right], 
\end{equation}
where we denote $\Delta \equiv 1/q$ and $\mathcal{J} \equiv J \sqrt{2^{1-1/\Delta}/\Delta}$.
The temperature and energy density are related via~\cite{maldacena2016remarks}
\begin{equation}
    e = e_0 + \frac{1}{2} c \beta^{-2}, \,\,\,\,\,\, c = 4\pi^2 \alpha_S / \mathcal{J},
\end{equation}
where $c$ is the specific heat, and $\alpha_S$ is a constant that approaches $\alpha_S \rightarrow 1/4q^2$ as $q \rightarrow \infty$.
Following the steps of the previous section, these relations lead to an effective Schrodinger equation, Eq.~(\ref{eq: QSHO}), with a potential minimum, spring constant, and mass,
\begin{equation} \label{eq: parameters SYK}
\beta^*_{\text{SYK}} = \left( \frac{\mu \Delta}{c} \cdot c_\Delta \left( \frac{\pi}{\mathcal{J}} \right)^{2\Delta} \right)^{\frac{-1}{2-2\Delta}},\,\,\,\,\,\,\,\, k_{\text{SYK}}(\mu) = \frac{4(1-\Delta)}{c} (\beta^*_{\text{SYK}})^2, \,\,\,\,\,\,\, m_{\text{SYK}}(\mu) = \frac{(\beta^*_{\text{SYK}})^4}{2\pi^2 c}.
\end{equation}
The inverse temperature at the potential minimum, $\beta^*_{\text{SYK}}$, agrees precisely with the predictions from the the low-energy gravitational theory (see Refs.~\cite{maldacena2018eternal,chen2019entanglement}; note that our inverse temperature is related to the variable $t'$ in these works via $(\beta^*_{\text{SYK}})^{-1} = \mathcal{J} t^\prime / 2 \pi \alpha_S$).
These harmonic oscillator parameters lead to a gap,
\begin{equation}
    \omega_{\text{SYK}} =  \frac{2\pi \sqrt{2(1-\Delta)}}{\beta^*_{\text{SYK}}},
\end{equation}
which again, agrees exactly with predictions for the graviton mode in the coupled SYK Hamiltonian~\cite{maldacena2018eternal}.

In what follows, it will be convenient to change variables in the effective Schrodinger equation, from the energy density $e$ to the logarithm of the inverse temperature, $\phi(e) \equiv \log( J \beta(e) )$.
This yields a Schrodinger equation,
\begin{equation} \label{eq: QSHO phi}
i N^{-1} \partial_t \psi(\phi) \approx \left( \text{constant} + \frac{ 1 }{2} \tilde{k}_{\text{SYK}}(\mu) (\phi-\phi^*_{\text{SYK}})^2 \right) \psi(\phi) - \frac{ N^{-2} }{2 \tilde{m}_{\text{SYK}}} \d_\phi^2 \psi(\phi),
\end{equation}
with a potential minimum, spring constant, and mass,
\begin{equation}
\phi^*_{\text{SYK}} = \frac{1}{2-2\Delta} \log\left( \frac{c}{\mu J \Delta} \cdot c_\Delta^{-1} \left( \frac{\pi}{\mathcal{J}} \right)^{-2\Delta} \right),\,\,\,\,\,\,\,\, \tilde{k}_{\text{SYK}}(\mu) = \frac{4(1-\Delta)c}{(\beta^*_{\text{SYK}})^2}, \,\,\,\,\,\,\, \tilde{m}_{\text{SYK}} =  \frac{c}{2\pi^2}.
\end{equation}
Compared to the previous coordinates, the spring constant and mass are both rescaled by $\left( \d e / \d \phi \right)^2 = c^2/\beta^4$.
As we will see in the later sections, the $\phi$ coordinate is convenient because the effective mass, $\tilde{m}_{\text{SYK}}$, becomes independent of $\mu$.
Thus, even when $\mu$ is time-dependent, we can still view the system as a particle moving in a one-dimensional (time-dependent) potential.
In particular, when the potential is moving to the right at a constant velocity, i.e.~$\phi^*_{\text{SYK}}$ is increasing at a constant linear rate in time, we can transform into an inertial frame in which the center of the potential is stationary.
This will substantially simplify our analysis of the adiabatic protocol in the following sections.

\subsection{Accuracy of the framework} \label{sec: accuracy}

%

We conclude this section by providing a basic self-consistency check to benchmark the accuracy of this theoretical framework.
Our check is to compute the variance of the coupled Hamiltonian in the putative ground state~\cite{fn6}.
If the variance is small, then the predicted ground state is close to an eigenstate of the coupled Hamiltonian, suggesting that the framework can be trusted.
If the variance is large, then the predicted ground state must differ from the true ground state, and so the framework should be treated much more cautiously.

To compute the energy variance, $V$, we expand the coupled Hamiltonian into its paired and unpaired components, $H = H_{\text{p}} + H_{\text{u}}$, where $H_{\text{p}}$ contains all matrix elements between states of the form $\ket{mm^*}$, $\ket{nn^*}$, and $H_{\text{u}}$ contains all other matrix elements.
Our theoretical framework analyzes solely $H_{\text{p}}$ and obtains the exact ground state, $\ket{\psi_{\text{p}}}$, of this component of the Hamiltonian.
Thus, the variance of the full Hamiltonian in $\ket{\psi_{\text{p}}}$ arises solely from the unpaired component,
\begin{equation}
    V = \bra{\psi_{\text{p}}} H^2 \ket{\psi_{\text{p}}} - \bra{\psi_{\text{p}}} H \ket{\psi_{\text{p}}}^2 = \bra{\psi_{\text{p}}} H_{\text{u}}^2 \ket{\psi_{\text{p}}}
    = \mu^2 \sum_{j,j'} \bra{\psi_{\text{p}}} (i\chi^j_L \chi^j_R)_{\text{u}} (i\chi^{j'}_L \chi^{j'}_R)_{\text{u}} \ket{\psi_{\text{p}}},
\end{equation}
where the subscript denotes that we are considering only the  matrix elements of the coupling that involve unpaired states.
To evaluate this expression, we first assume that to leading order in $1/N$, the expectation values above are zero unless $j=j'$.
This gives,
\begin{equation}
    V = \mu^2 \sum_{j} \bra{\psi_{\text{p}}} (i\chi^j_L \chi^j_R)_{\text{u}} (i\chi^{j}_L \chi^{j}_R)_{\text{u}} \ket{\psi_{\text{p}}}.
\end{equation}
We can evaluate the remaining expectation values by writing the unpaired component of the coupling as the difference between the full coupling and its paired component,
\begin{equation}
    \mu^2 \sum_{j} \bra{\psi_{\text{p}}} (i\chi^j_L \chi^j_R)_{\text{u}} (i\chi^{j}_L \chi^{j}_R)_{\text{u}} \ket{\psi_{\text{p}}}
    =
    \mu^2 \sum_{j} \bra{\psi_{\text{p}}} [ i\chi^j_L \chi^j_R - (i\chi^j_L \chi^j_R)_{\text{p}} ] [ i\chi^{j}_L \chi^{j}_R - (i\chi^{j}_L \chi^{j}_R)_{\text{p}} ] \ket{\psi_{\text{p}}}.
\end{equation}
The expression on the right contains four cross terms.
The first is,
\begin{equation}
    \mu^2 \sum_j \bra{\psi_{\text{p}}} [ i\chi^j_L \chi^j_R ] [ i\chi^{j}_L \chi^{j}_R ] \ket{\psi_{\text{p}}} =  \mu^2 \sum_j 1 = \mu^2 N,
\end{equation}
where each expectation value evaluates to one because $[i(\chi^j_L \chi^j_R)]^2 = \mathbbm{1}$.
The latter three cross terms contain expectation values of the form,
\begin{equation}
    \bra{\psi_{\text{p}}} i\chi^j_L \chi^j_R (i\chi^{j}_L \chi^{j}_R)_{\text{p}} \ket{\psi_{\text{p}}} = \bra{\psi_{\text{p}}} (i\chi^j_L \chi^j_R)_{\text{p}} (i\chi^{j}_L \chi^{j}_R)_{\text{p}} \ket{\psi_{\text{p}}}.
\end{equation}
We can re-write these as the sum of two terms,
\begin{equation}
    \bra{\psi_{\text{p}}} (i\chi^j_L \chi^j_R)_{\text{p}} (i\chi^{j}_L \chi^{j}_R)_{\text{p}} \ket{\psi_{\text{p}}} = \bra{\psi_{\text{p}}} (i\chi^{j}_L \chi^{j}_R)_{\text{p}} \ket{\psi_{\text{p}}}^2 + V' = G_{\beta^*}^2 + V',
\end{equation}
where $V'$ is the variance of the operator $(i\chi^j_L \chi^j_R)_{\text{p}}$ in $\ket{\psi_{\text{p}}}$.
In the final expression, we have used the fact that the expectation value of the coupling in the ground state is given by the two-point function, $G_{\beta^*}$, where $\beta^*$ is the inverse temperature of the associated thermofield double state.
From the previous sections, the variance $V'$ is suppressed by $1/N$ since the ground state has width $1/\sqrt{N}$ in the energy density.
Thus, we are free to drop the $V'$ term in the large-$N$ limit.

All told, the above analysis yields a variance,
\begin{equation}
    V = \mu^2 ( 1 - G_{\beta^*}^2 ) N,
\end{equation}
which indicates the existence of energy fluctuations of order $\mu (1-G_{\beta^*}^2)^{1/2}$ per fermion.
To quantify the accuracy of the framework, we should compare the magnitude of these fluctuations to the local energy scale, ${\beta^*}^{-1}$.
In particular, by taking the ratio of the energy variance per fermion with the squared temperature, we obtain a dimensionless number,
\begin{equation} \label{eq: dimensionless}
    \frac{V}{N ({\beta^*})^{-2}} = (\beta^* \mu)^2 ( 1 - G_{\beta^*}^2 ).
\end{equation}
This quantifies the relative local error in the framework. We recall that the value of $\beta^*$ is set by $\mu$ (Section~\ref{sec: SYK ETH}).

Let us now analyze the magnitude of the error as a function of the effective temperature.
For brevity, we now denote the inverse effective temperature as $\beta$ and not $\beta^*$.
At low temperatures, $\beta J \gg 1$, the first term in Eq.~(\ref{eq: dimensionless}) is small since $\mu = \mathcal{O}((\beta J)^{-2+2\Delta})$ [Eq.~(\ref{eq: parameters SYK})]. Thus, the relative error is also small,
\begin{equation}
    \,\,\,\,\,\,\,\,\,\,\,\,\,\,\,\,\,\,\,\,\,\, \frac{V}{N \beta^{-2}} = \mathcal{O}((\beta J)^{-2+4\Delta}), \,\,\,\,\,\,\,\,\,\,\, \text{at low temperatures, } \beta J \gg 1,
\end{equation}
and the framework is valid.
This was already somewhat expected, since the effective Schrodinger equation in Section~\ref{sec: SYK ETH} matches exactly with the known graviton mode of the SYK model at low temperatures~\cite{maldacena2018eternal}.
A similar relative error was derived in Ref.~\cite{maldacena2018eternal}, via different means, to confirm the validity of their analysis.

At high temperatures, the first term in Eq.~(\ref{eq: dimensionless}) is $\mathcal{O}(1)$ and, instead, the \emph{second} term is small, $\mathcal{O}((\beta J)^2)$, since $G_\beta$ approaches one.
This leads to a relative error,
\begin{equation} \label{eq: relative error high}
    \,\,\,\,\,\,\,\,\,\,\,\,\,\,\,\,\,\,\,\,\,\, \frac{V}{N \beta^{-2}} = \mathcal{O}((\beta J)^2), \,\,\,\,\,\,\,\,\,\,\, \text{at high temperatures, } \beta J \ll 1,
\end{equation}
which demonstrates that our framework is also valid at high temperatures.
Notably, this holds for any high temperature, $\beta J \ll 1$, and so is significantly more general than the specific  $\mu = \infty$ limit in which the coupled Hamiltonian reduces to a sum of EPR projectors.
To derive Eq.~(\ref{eq: relative error high}), we first note that at high temperatures the two-point function will generically decay quadratically with the inverse temperature, $G_\beta \approx 1 - a (\beta J)^2$, for some $a = \mathcal{O}(1)$~\cite{maldacena2016remarks}.
Thus, the second term in Eq.~(\ref{eq: dimensionless}) is $\mathcal{O}((\beta J)^2)$ as claimed.
To determine the first term, we recall that $\mu$ is related to the inverse temperature via $\mu = 2 / \partial_e G_{\beta(e)} = 2 (d \beta/d e) / \partial_\beta G_\beta$ [Eq.~(\ref{eq: set e min})].
At high temperatures, the energy density will typically decrease linearly in the inverse temperature, $e \sim - b J^2 \beta$, for some $b = \mathcal{O}(1)$~\cite{garcia2017analytical}.
This gives $d \beta / d e = - 1/bJ^2$. We also have $\partial_\beta G_\beta = -2a\beta J^2$ from the functional form of $G_\beta$ above.
Together, these yield $\mu = \mathcal{O}(\beta^{-1})$, and thus the first term in Eq.~(\ref{eq: dimensionless}) is indeed $\mathcal{O}(1)$.

At intermediate temperatures, $\beta J \sim 1$, the relative error is generically $\mathcal{O}(1)$ and so the framework is not guaranteed to be valid.
Nonetheless, as discussed in the main text, we find a surprisingly good quantitative agreement between the framework's predictions and our large-$N$ numerics at all temperatures.
This suggests that the framework continues to capture much of the essential physics of the coupled Hamiltonian even at intermediate temperatures.

\section{Analysis of quantum adiabatic protocol}

In the main text, we outlined how the quantum mechanical adiabatic theorem enables one to prepare the ground state of the coupled SYK Hamiltonian with high many-body fidelity, by slowly interpolating the coupling strength from a large to small value.
Here, we provide a detailed accounting of the errors in the adiabatic approximation.
This will allow us to estimate the evolution time required by our cooling protocol to ensure that all error sources are small.
For simplicity, we work in the low-temperature limit and focus on the scaling of the required evolution time with desired inverse temperature, $\beta$.

As introduced in the main text, we consider the following exponential ramp in the coupling, $\mu(t)$, from a high value, $\mu_i$, to a low value, $\mu_f$, over a time $T$,
\begin{equation} \label{eq: exponential ramp mu}
    \mu(t) = \mu_i e^{-(t/T) \log(\mu_i/\mu_f)}.
\end{equation}
For small coupling strengths, the inverse temperature of the associated  thermofield double state is given by, $\beta^*(\mu) \sim \mu^{2/3}$ (see Section~\ref{sec: SYK ETH}).
Thus, under this ramp, $\beta^*$ increases exponentially in time, $\beta^*(t) = \beta^*_i e^{(t/T) \text{log}(\beta^*_i/\beta^*_f)}$, with $\beta^*_{i,f}$ set by $\mu_{i,f}$.
As discussed in Section~\ref{sec: SYK ETH}, we will find it more convenient to work with the logarithm of the inverse temperature, $\phi^* = \text{log} (\beta^* J)$.
The exponential ramp in $\beta^*$ corresponds to a simple linear ramp in $\phi^*$,
\begin{equation} \label{eq: linear ramp phi}
    \phi^*(t) = \phi^*_i + \frac{t}{T} \Delta \phi,
\end{equation}
where $\phi^*_i = \log(\beta^*_i J)$ and $\Delta \phi = \text{log}( \beta^*_f / \beta^*_i )$.
To analyze the scaling of $T$ with $\beta^*_f$, in what follows, we will fix $\beta^*_i = \mathcal{O}(J^{-1})$ to be a small constant multiplied by $J$, and denote the desired inverse temperature $\beta^*_f \equiv \beta$ for brevity.

At a high level, errors in the adiabatic approximation will fall under two types, corresponding to the classes of excitations in the coupled SYK Hamiltonian.
The first type of errors are associated with matter excitations, i.e.~``unpaired'' excited states in the framework of Section~\ref{app: ETH}.
We will find that the magnitude these errors can be estimated via a standard adiabatic treatment.
The latter type of errors are associated with graviton excitations, i.e.~``paired'' excited states in Section~\ref{app: ETH}.
These errors will require a more nuanced treatment, due an emergent Galilean invariance in the effective Schrodinger equation that governs the graviton dynamics.
In particular, we will identify three distinct sources of graviton excitations in the adiabatic protocol.

In more detail, we find the following four sources of potential errors to the adiabatic protocol, whose magnitudes and physical origins are summarized below. 

\begin{enumerate}
    \item \textbf{Adiabatic leakage to matter excitations:} As we lower $\mu$, the time-dependent Hamiltonian generates a small amount of matter excitations. 
    These arise due to a non-zero overlap of the $\mu$-derivative of the ground state, $\sim \! (\sum_j i \chi^j_L \chi^j_R) \ket{\psi_{\text{gs}}}$, with excited states that contain a matter excitation both sides of the thermofield double.
    We estimate that such excitations contribute an infidelity $\varepsilon_m = \mathcal{O}(N \log(\beta J) / J T)$ to the adiabatic protocol.

    \item \textbf{Sudden start and stop of the ramp:} The instantaneous acceleration and deceleration of $\phi^*$ at times zero and $T$ generates a small amount of graviton excitations. By transforming to a moving reference frame in the effective Schrodinger equation, we estimate that such excitations contribute an infidelity, $\varepsilon_a = \mathcal{O}(N \, \beta  \log^2(\beta J) / J T^2)$.

    \item \textbf{Adiabatic change in the spring constant:} The spring constant $\tilde{k}(\mu)$ of the effective Schrodinger equation decreases exponentially in time as $\phi^*$ is increased. This causes the ground state wavefunction (when viewed as a function of $\phi^*$) to broaden over time. A standard treatment shows that corrections to adiabaticity due to this broadening contribute a small infidelity $\varepsilon_{sc} = \mathcal{O}(\beta \log(\beta J) / T)$, independent of the system size.

    \item \textbf{Robustness to perturbations:} If the ramp in $\phi^*$ is perfectly linear in time, then, in the low-temperature limit, the motion of the potential minimum \emph{during} the ramp generates no excitations. 
    This surprising fact is due to an emergent Galilean symmetry in the effective Schrodinger equation, Eqs.~(\ref{eq: eff H main text},~\ref{eq: QSHO phi}), which allows one to transform to a moving frame in which the potential minimum is stationary.
    However, this emergent symmetry could in principle be broken, for example due to imprecisions in the ramp rate, or higher-temperature corrections to the graviton mode.
    To demonstrate that our estimated evolution time is not reliant on this emergent symmetry, we consider the performance of the adiabatic protocol when the ramp in $\phi^*$ is not perfectly linear in time.
    We estimate that imperfections of magnitude $\delta = \delta \phi^* / \phi^*$ will generate additional graviton excitations throughout the ramp, with an associated infidelity $\varepsilon_{p} = \mathcal{O}( N \delta^2 \log^2( \beta J) / J T)$.
    
\end{enumerate}
\noindent The first and last error source provide the leading contributions for large $N$ and $T$. 
Focusing on the first, since the parameter $\delta$ in the last error source can in principle be suppressed, we find a required evolution time, $T = \Omega( N \log(\beta J)/J)$, to prepare the ground state of the coupled Hamiltonian to high many-body fidelity.

In the following subsections, we provide detailed discussions and derivations of each of the error estimates above.
For the interested reader, in a final subsection, we provide additional analyses for other, non-exponential time-profiles of the ramp in $\beta^*(t)$.
We find that both linear and inverse linear ramps in the temperature lead to sub-optimal preparation times.

\subsection{Adiabatic leakage to matter excitations} \label{sec: qu matter}

We begin our analysis by considering leakage to matter excitations, i.e.~``unpaired'' excited states in Section~\ref{sec: SYK ETH}.
In the adiabatic limit, we can form a Fermi's golden rule estimate of the total leakage, $\varepsilon_m$, to matter excitations over time,
\begin{equation}
    \varepsilon_m \approx N \int_0^T dt \, \frac{| \langle \psi_{\text{e},m} | \partial_t | \psi_{\text{gs}} \rangle |^2}{\Delta_m} = N \int_0^T d{\phi^*} \, \left( \frac{\d \phi^*}{\d t} \right) \frac{| \langle \psi_{\text{e},m} | \partial_\phi^* | \psi_{\text{gs}} \rangle |^2}{\Delta_m},
\end{equation}
where $\ket{\psi_{\text{e},m}}$ is the eigenstate of a single matter excitation, $\Delta_m = \mathcal{O}(\beta^*)$ is the gap to matter excitations, and the factor of $N$ accounts for the fact that there are $N$ degenerate such excited states (corresponding to each Majorana operator $\chi_i$ in the SYK model).
On the RHS, we switch coordinates from $t$ to $\phi^*$.

We can estimate the matrix element in the above equation as follows.
First, we take the following rough ansatz for the lowest-lying excited state that is coupled to the ground state by the Hamiltonian,
\begin{equation}
    \ket{\psi_{\text{e},m}} \approx \frac{ \big( i \chi^j_L \chi^j_R \big)_{\text{u}}  \ket{ \psi_{\text{gs}} } }{ \sqrt{ \bra{ \psi_{\text{gs}} } \big( i \chi^j_L \chi^j_R \big)_{\text{u}}^2  \ket{ \psi_{\text{gs}} } } },
\end{equation}
where $j$ is arbitrary and $\big( i \chi^j_L \chi^j_R \big)_{\text{u}}$ denotes the component of the operator $i \chi^j_L \chi^j_R$ that couples paired to unpaired states, in the ETH parlance.
%
%
The state written is clearly not an exact eigenstate of the Hamiltonian, but we anticipate that it has large overlap with such a state.
Moreover, the denominator that normalizes the state is order one,
\begin{equation}
    \bra{ \psi_{\text{gs}} } \big( i \chi^j_L \chi^j_R \big)_{\text{u}}^2  \ket{ \psi_{\text{gs}} } = \bra{ \psi_{\text{gs}} } \big( i \chi^j_L \chi^j_R - ( i \chi^j_L \chi^j_R )_{\text{p}} \big)^2  \ket{ \psi_{\text{gs}} } = 1 - G_{\beta^*} \sim \mathcal{O}(1),
\end{equation}
so we will neglect it from hereon.
The desired matrix element is
\begin{equation}
    \langle \psi_{\text{e},m} | \partial_\phi^* | \psi_{\text{gs}} \rangle \sim \langle \psi_{\text{gs}} | \big( i \chi^j_L \chi^j_R \big)_{\text{u}} \partial_\phi^* | \psi_{\text{gs}} \rangle.
\end{equation}
To compute it, we follow the calculations of Ref.~\cite{maldacena2018eternal} and write
\begin{equation}
\begin{split}
    | \psi_{\text{gs}} \rangle = \lim_{\tau \rightarrow \infty} \frac{1}{\mathcal{N}_\tau^{1/2}}  e^{-\tau H} \ket{\text{EPR}},
\end{split}
\end{equation}
with $H = H_L + H_R + \mu \sum_j i \chi^j_L \chi^j_R$, and $\mathcal{N}_\tau$ defined to normalize the state.
We can now take the derivative over $\mu$ explicitly, yielding,
\begin{equation}
\begin{split}
    \langle \psi_{\text{gs}} | \big( i \chi^j_L \chi^j_R \big)_{\text{u}} \partial_\mu | \psi_{\text{gs}} \rangle & = \lim_{\tau \rightarrow \infty} \frac{1}{\mathcal{N}_\tau^{1/2}} \langle \psi_{\text{gs}} | \big( i \chi^j_L \chi^j_R \big)_{\text{u}} \cdot \int_0^\tau ds \, e^{-s H} \big( \sum_j i \chi^j_L \chi^j_R  \big) e^{-(\tau-s)H}  | \text{EPR} \rangle \\
    & = \int_0^\infty ds \, \langle \psi_{\text{gs}} | \big( i \chi^j_L \chi^j_R \big)_{\text{u}} \cdot \big( i \chi^j_L(s) \chi^j_R(s) \big) | \psi_{\text{gs}} \rangle, \\
\end{split}
\end{equation}
where the latter pair of operators are imaginary time-evolved by $s$.
The final inner product can be evaluated by Wick contracting pairs of the four fermions.
Taking the unpaired component of the first pair of fermions amounts to excluding the Wick contraction between that pair.
The dominant remaining Wick contractions correspond to pairing the two $\chi^j_L$ fermions with each other, and similar for the $\chi^j_R$ fermions.
Each contraction gives a factor of $\bra{ \psi_{\text{gs}} } \chi_L^j(0) \chi_L^j(s) \ket{ \psi_{\text{gs}} } \sim \left( \beta^* J / \sinh(\beta^* s) \right)^{2\Delta}$~\cite{maldacena2018eternal}, yielding an overlap
\begin{equation}
\begin{split}
    \langle \psi_{\text{gs}} | \big( i \chi^j_L \chi^j_R \big)_{\text{u}} \partial_\mu | \psi_{\text{gs}} \rangle & = \int_0^\infty ds \, \left( \frac{ (\beta^* J)^{-1} }{ \sinh(s / \beta^* ) } \right)^{4\Delta} \sim \mathcal{O}\left( \beta^* (\beta^* J)^{-4\Delta} \right).
\end{split}
\end{equation}
Changing coordinates from $\mu$ to $\phi^*$ via $\d \mu / \d \phi^* = \mathcal{O}(\mu) = \mathcal{O}( J (\beta^* J)^{2\Delta -2} )$, we have
\begin{equation}
    \langle \psi_{\text{gs}} | \big( i \chi^j_L \chi^j_R \big)_{\text{u}} \partial_\phi^* | \psi_{\text{gs}} \rangle = \mathcal{O}\left( \beta^* (\beta^* J)^{-4\Delta} J (\beta^* J)^{2\Delta-2} \right) = \mathcal{O}\left( \frac{1}{(\beta^* J)^{2\Delta+1}} \right).
\end{equation}

We can now, finally, complete our estimation of the leakage to matter excited states.
We have
\begin{equation} \label{eq: error matter}
    \varepsilon_m \sim N \cdot \frac{\log(\beta_i^*/\beta)}{T} \int d \phi^* \, \frac{(\beta^* J)^{-4\Delta-2}}{1/\beta^*} \sim N \frac{\log(\beta J)}{J T} \int d \phi^* \, e^{-(4\Delta+1)\phi^*} = \mathcal{O}\left( \frac{N \log( \beta J) }{JT} \right).
\end{equation}
We see that the leakage to matter excited states occurs primarily at early times in the protocol,~i.e.~small $\phi^*$ in the integral above.
This is due to the rapid decrease in the matrix element as the temperature is lowered, which overwhelms the much slower decrease in the excitation gap.
Thus, the total leakage is small whenever $T > \mathcal{O}(N \log(\beta J)/J)$.

\subsection{Sudden start and stop of the ramp}

We now discuss the leakage to graviton excitations, i.e.~``paired'' excited states.
From Section~\ref{sec: SYK ETH}, we know that the dynamics of the graviton sector are governed by a simple one-dimensional harmonic oscillator, where the coordinate $\phi$ plays the role of the position axis.
The exponential decrease in $\mu$ corresponds to a simple linear ramp in potential minimum of the harmonic oscillator, $\phi^*(\mu)$.
Over the course of this ramp, the spring constant, $\tilde{k}(\mu)$, decreases exponentially as a function of time, while the mass, $\tilde{m}$, remains constant.

Since the potential minimum $\phi^*(t)$ increases linearly in time, during the ramp, we are free to transform to a moving frame, with velocity $\partial \phi^* / \partial t$, in which the potential minimum is stationary.
This moving frame is an inertial frame since the mass is independent of the position and time.
Thus, in this frame, the population of the system in the ground and excited states remains constant over the course of the ramp (neglecting non-adiabatic effects due to the change in $\tilde{k}(\mu)$ over time, which we address in the following section).
This means that graviton excitations are incurred only at the very start and stop of the ramp, when the potential minimum is instantaneously accerates or decelerates.

To estimate the magnitude of these excitations, we assume the system begins in the ground state of the low-energy effective Schrodinger equation at time zero.
Just after time zero, the potential minimum begins moving at a constant rate.
In the moving frame, the original, stationary ground state has wavefunction
\begin{equation}
    \tilde{\psi}_{\text{gs}}(\phi) = \left( \frac{N \tilde{m} \omega_i}{\pi} \right)^{1/4} \exp( - \frac{N \tilde{m}\omega_i}{2} ( \phi - \phi^*_i)^2 ) \exp( -i N \tilde{m} v \phi ),
\end{equation}
where $\omega_i$ is the frequency at time zero, $v \equiv \partial \phi^*/\partial t$ is the velocity of the potential minimum, and $\tilde{m} v$ is the associated momentum.
The latter exponential implements the boost, $\exp(i p x / \hbar )$, from the stationary to rotating frame. 
The infidelity associated with this boost is given by one minus the overlap of $\tilde{\psi}_{\text{gs}}$ with the ground state, $\psi_{\text{gs}}$, in the moving frame.
This is simply the expectation value of $\exp(-i N \tilde{m} v \phi)$, which gives
\begin{equation}
    \varepsilon_{m,i} = 1 - \exp( - \frac{1}{2} \frac{\tilde{m} v^2 / 2}{\omega_i/ N} ).
\end{equation}
The argument of the exponential is ratio of the kinetic energy in the rotating frame with the energy gap of the harmonic oscillator (in units of $N$).
Noting that $v = \Delta \phi / T = \log(\beta J) / T$, $\tilde{m} = \mathcal{O}(J^{-1})$, and $\omega_i^{-1} = \mathcal{O}(\beta^*_i) = \mathcal{O}(J^{-1})$ we have $\varepsilon_{a,i} = \mathcal{O}( N \log^2(\beta J) / J^2 T^2)$.
Performing an identical analysis for the \emph{stop} of the ramp at time $T$, we find the same infidelity with $\omega_i = \mathcal{O}(J)$ replaced by $\omega_f = \mathcal{O}(\beta^{-1})$, which yields $\varepsilon_{a,f} = \mathcal{O}( N \beta \log^2(\beta J) / J T^2)$.
The total infidelity, $\varepsilon_a  \approx \varepsilon_{a,i} + \varepsilon_{a,f}$, is dominated by the stopping term owing to the smaller energy gap.

\subsection{Adiabatic change in the spring constant}

While we found in the previous section that the \emph{motion} of the potential minimum did not generate any excitations during the ramp (due to the emergent Galilean invariance), the change in the spring constant $\tilde{k}(\mu)$ during the ramp \emph{can} lead to excitations.
To derive the infidelity associated with these excitations, we apply the standard Fermi's golden rule estimate,
\begin{equation}
    \varepsilon_{sc} \approx \int_0^T dt \left(\frac{\partial \tilde{k}}{\partial t} \right)^2 \frac{| \bra{\psi_{\text{e}}} \partial_{\tilde{k}} \ket{\psi_{\text{gs}}} |^2}{\omega(t)} = \frac{\Delta \phi}{T} \int d\phi^* \left(\frac{\partial \tilde{k}}{\partial \phi^*} \right)^2 \frac{| \bra{\psi_{\text{e}}} \partial_{\tilde{k}} \ket{\psi_{\text{gs}}} |^2}{\omega(\phi^*)},
\end{equation}
where $\omega(t)$ is the gap of the harmonic oscillator at time $t$ during the ramp, and $\ket{\psi_{\text{e},g}}$ is the second excited state of the harmonic oscillator (the matrix element to the first excited state is zero by symmetry arguments).
By dimensional analysis, we have $\partial \tilde{k} / \partial_{\phi^*} = \mathcal{O}(\tilde{k})$ and $| \bra{\psi_{\text{e}}} \partial_{\tilde{k}} \ket{\psi_{\text{gs}}} | = \mathcal{O}(\tilde{k}^{-1})$; both of these can also be verified explicitly.
The factors of $\tilde{k}$ cancel, and we find an infidelity
\begin{equation}
    \varepsilon_{sc} \sim \frac{\Delta \phi}{T} \int d \phi^* \frac{1}{\omega(\phi^*)} = \mathcal{O}\left( \frac{\Delta \phi}{T} \cdot \omega_f \right) =  \mathcal{O}\left( \frac{\beta \log( \beta J)}{T} \right).
\end{equation}
The integral is dominated by late times where the gap, $\omega_f = \mathcal{O}(\beta^{-1})$, is small.
The infidelity does not scale with the system size $N$, because the matrix element is independent of $N$.

\subsection{Robustness to perturbations}

Finally, we analyze the robustness of the quantum adiabatic protocol to perturbations that break the emergent Galilean symmetry of the effective Schrodinger equation.
For example, these might arise due to imperfections in a quantum experiment that cause the ramp rate to not be perfectly linear, or due to corrections to the effective Schrodinger equation that are suppressed at low temperatures.
We will show, via a straightforward adiabatic analysis, that the required evolution time in the previous sections is robust to such perturbations.

As mentioned in the introduction, we consider the simplest possible source of broken Galilean symmetry: imperfections in the time-profile of the ramp, $\phi^*(t)$.
To capture such imperfections, let $\delta \phi^*(t) = \tilde{\phi}^*(t) - \phi^*(t)$ denote the difference between the actual ramp, $\tilde{\phi}^*(t)$, and the ideal ramp, $\phi^*(t)$.
We assume that $\delta \phi(t)$ varies slowly along the ramp, $| \partial \delta \phi / \partial \phi^*| \leq \delta$, so that the evolution remains adiabatic, but otherwise $\delta \phi^*(t)$ may have an arbitrary functional form.
We work in the moving frame of the ideal ramp, so that the true potential minimum lies at $\delta \phi^*(t)$.

We can estimate the leakage to excited states via the standard Fermi's golden rule, 
\begin{equation}
    \varepsilon_p \sim \int dt \left( \frac{\partial \delta \phi^*}{\partial t} \right)^2 \frac{| \bra{ \psi_{\text{e}} } \partial_{\delta \phi^*} \ket{ \psi_{\text{gs}}} |^2}{\omega(t)} = \frac{\Delta \phi}{T} \int d \phi^* \left( \frac{\partial \delta \phi^*}{\partial \phi^*} \right)^2 \frac{| \bra{ \psi_{\text{e}} } \partial_{\delta \phi^*} \ket{ \psi_{\text{gs}}} |^2}{\omega(\phi^*)}.
\end{equation}
By assumption, the first term in the second integral is less than $\delta^2$.
Meanwhile, the matrix element is given the inverse radius of the harmonic oscillator,
\begin{equation}
    | \bra{ \psi_{\text{e}} } \partial_{\delta \phi^*} \ket{ \psi_{\text{gs}}} | \approx \sqrt{ N \tilde{m} \omega(\phi^*)}.
\end{equation}
Taking the square of the matrix element and dividing by the gap, we find that all time-dependence in the integrand cancels, yielding
\begin{equation} \label{eq: error imprecision}
    \varepsilon_p \sim  \frac{\Delta \phi}{T} \int d \phi^* \delta^2  N \tilde{m} = \mathcal{O}\left( \frac{N \delta^2 \log^2(\beta J) }{J T} \right).
\end{equation}
The scaling is identical to that of the infidelity due to matter excitations, aside from an additional logarithmic factor.
Thus, as long as the ramp in $\phi^*$ is close to linear, we will achieve the stated evolution time.

\subsection{Other ramp profiles}

We can also easily extend our analysis to different time-profiles of the ramp.
We consider two such profiles: a linear ramp in the inverse temperature, $\beta^*(t)$, and a linear ramp in temperature, $(\beta^*)^{-1}(t)$.
We will see that both examples give poorer scalings of the error compared to the exponential ramp.
For simplicity, we analyze only the first and last error source, since these give the leading order behavior in $N$.

For the linear ramp in the inverse temperature, we take $\beta^*(t) = \beta^*_i + (t/T) \beta^*_f$.
To analyze the effect of this choice of $\beta^*(t)$ on the infidelity, we need only to compute the time-derivative of $\phi^*(t) = \log(\beta^*(t) J)$.
We have $\partial \phi^* / \partial t = (1/T) (\beta_f^* - \beta_i^*) / \beta^* = \mathcal{O}( (1/T) (\beta/\beta^*))$.
Thus, we simply need to replace the factors of $\Delta \phi$ in the previous error estimates with the new, time-dependent factor $\beta/\beta^*(t)$.
In Eq.~(\ref{eq: error matter}), this leads to an infidelity
\begin{equation}
    \varepsilon_m \sim \frac{N}{T} \int_0^T d \phi^* \, \frac{\beta}{\beta^*} \frac{(\beta^* J)^{-4\Delta-2}}{1/\beta^*}  = \mathcal{O}\left( \frac{N }{\beta^{-1} T} \right),
\end{equation}
which is worse than that of the exponential ramp by a factor of $\beta J$.
In Eq.~(\ref{eq: error imprecision}), this leads to an infidelity
\begin{equation} 
    \varepsilon_p \sim  \frac{1}{T} \int d \phi^* \frac{\beta}{\beta^*} \delta^2  N \tilde{m} = \mathcal{O}\left( \frac{N \delta^2 }{\beta^{-1} T} \right),
\end{equation}
which is similarly worse.

Turning the linear ramp in the \emph{temperature}, we take $(\beta^*)^{-1}(t) = (\beta^*_i)^{-1} + (t/T) (\beta^*_f)^{-1}$.
We now have $\partial \phi^* / \partial t = (1/T) ((\beta_f^*)^{-1} - (\beta_i^*)^{-1})/ (\beta^*)^{-1} = \mathcal{O}( (1/T) J \beta^*)$.
In Eq.~(\ref{eq: error matter}), this leads to an infidelity 
\begin{equation}
    \varepsilon_m \sim \frac{N}{T} \int_0^T d \phi^* J \beta^* \cdot \frac{(\beta^* J)^{-4\Delta-2}}{1/\beta^*}  = \mathcal{O}\left( \frac{N }{J T} \right),
\end{equation}
which is very mildly improved compared to the original scaling.
The improvement arises because the new ramp is slower than the exponential ramp at high $\mu$, where the matrix elements to matter excited states are largest.
In Eq.~(\ref{eq: error imprecision}), this leads to an infidelity
\begin{equation} 
    \varepsilon_p \sim  \frac{1}{T} \int d \phi^* J \beta^* \delta^2  N \tilde{m} = \mathcal{O}\left( \frac{N \delta^2 }{\beta^{-1} T} \right),
\end{equation}
which is worse than the exponential ramp by a factor of $\beta J$.

\section{Analysis of semi-classical adiabatic protocol}

Let us now turn to the second variant of our cooling protocol, in which we decrease the coupling $\mu$ at a much faster rate in time.
Specifically, we again consider an exponential ramp in $\mu(t)$, as in Eq.~(\ref{eq: exponential ramp mu}), but we now take $T$ to be independent of the number of fermions $N$.
%
%
In contrast to the first, slower variant of the cooling protocol, this rapid ramp in $\mu$ will cause the system to almost immediately gain population entirely in excited states of the coupled Hamiltonian.
Thus, the efficacy of this second variant of the protocol relies more crucially on the specific excited state dynamics of the coupled SYK Hamiltonian.
In what follows, we analyze these dynamics in detail and show that they enable cooling down to small constant temperatures, $\beta^{-1}$, in time $T = \mathcal{O}(\beta^{-1})$.
Compared to the first variant of the protocol, in this setting the protocol will not produce the precise ground state of the coupled Hamiltonian.
Instead, we obtain a state that is close to the ground state with respect to local observables, such as the energy density and effective temperature with respect to the individual uncoupled SYK Hamiltonians.

As discussed in the main text, we can understand the efficacy of the second variant of our cooling protocol by again observing the effective Schrodinger equation, Eqs.~(\ref{eq: eff H main text},\ref{eq: QSHO phi}).
As mentioned in earlier sections, this equation captures the ground state and graviton excited states of the coupled SYK Hamiltonian.
Crucially, when the action of the system is large compared to the effective Planck constant, $1/N$, the Schrodinger equation features a semi-classical limit. 
(Indeed, in Ref.~\cite{maldacena2018eternal}, the equation was originally derived in this limit, and quantized after.)
This is precisely the limit relevant for the second variant of our adiabatic protocol, in which the system rapidly excites to states with large action, $\mathcal{O}(1)$, i.e.~high quantum numbers, $\mathcal{O}(N)$.
In this limit, the dynamics of the coupled Hamiltonian are described by a simple \emph{classical} harmonic oscillator,
\begin{equation} \label{eq: classical SHO}
    H = \frac{1}{2 \tilde{m}} p^2 + \frac{1}{2} \tilde{k}(t) \left[ \phi - \phi^*(t) \right]^2,
\end{equation}
with frequency, $\omega = \mathcal{O}(1/\beta^*(t))$.
Here $p$ is the canonical conjugate of the coordinate $\phi$.
Our cooling protocol slowly translates the center of the harmonic oscillator from one location, $\phi^*_i$, to another, $\phi^*_f$.
If this translation is performed slowly, the oscillator will remain nearby the center throughout the protocol, up to small deviations, $\delta \phi$.
This corresponds to the energy density of the system (with respect to the individual uncoupled SYK Hamiltonians) remaining nearby the ground state value, up to small deviations\footnote{Here, we use $e$ to denote the excess energy density of the uncoupled SYK Hamiltonians above the uncoupled SYK ground state energy.}, $\delta e / e  = \mathcal{O}(\delta \phi)$.
These deviations translate to large quantum numbers, $\mathcal{O}(N \delta \phi)$, of the resulting state, but only small variations in local observables compared to their ground state values.

In what follows, we analyze this picture in more detail and quantify the various error sources that lead to non-zero energy fluctations, $\delta e$, in the final state of the protocol.
We summarize our results below, and provide full details in the following subsections.

\begin{enumerate}
    \item \textbf{Finite density of matter excitations:} As we lower $\mu$ at an $\mathcal{O}(1)$ rate with respect to $N$, the time-dependent Hamiltonian generates a small \emph{density} of matter excitations. 
    Paralleling our analysis in Section~\ref{sec: qu matter}, we find that this density is given by $n_m = \mathcal{O}(\log(\beta J)/JT)$.
    As each individual matter excitation has energy $\mathcal{O}(\beta^{-1})$, this density leads to an increase in the energy density, $\delta e_m = \mathcal{O}(\log(\beta J)/\beta J T)$.
    Comparing to the thermal energy, $e = \mathcal{O}(1/\beta^2 J)$, we have $\delta e_m / e = \mathcal{O}(\log(\beta J)/\beta^{-1} T)$.

    \item \textbf{Sudden start and stop of the ramp:} The instantaneous acceleration and deceleration of $\phi^*$ impart a non-zero kinetic energy, $\delta e_a = \mathcal{O}(\log^2(\beta J)/J T^2)$, to the classical ``particle''. 
    This leads to a relative increase in energy density, $\delta e_a / e = \mathcal{O}(\log^2(\beta J)/\beta^{-2} T^2)$.

    \item \textbf{Adiabatic change in the spring constant:} As mentioned above, the initial acceleration of $\phi^*$ imparts a small kinetic energy to the classical oscillator.
    Over the course of the ramp, the spring constant of the oscillator changes, which can, potentially, either amplify or de-amplify this energy.
    In order determine how the energy changes during the ramp, we utilize the notion of \emph{classical} adiabatic invariants.
    For the harmonic oscillator, the action, $J \equiv \delta e / \omega$, is an adiabatic invariant, and is thus conserved in time as long as the Hamiltonian parameters change slowly enough.
    To quantify how slow is needed, we estimate next-order corrections to classical adiabaticity and find that they are suppressed by a relative magnitude $\mathcal{O}(\log(\beta J) / \beta^{-1} T)$.

    \item \textbf{Robustness to perturbations:} Finally, we demonstrate that the evolution times estimated from the above errors are robust to perturbations that break the emergent Galilean invariance in the harmonic oscillator.
    As mentioned previously, such perturbations might arise from experimental imperfections in the ramp rate, or higher-temperature corrections to the graviton dynamics.
    As a toy model of such a perturbation, we consider the effect of adding a small frictional damping, of strength $\gamma$, to the harmonic oscillator dynamics.
    We find that such a damping leads to a relative increase, $\delta e_p / e = \mathcal{O}(\gamma \log^2(\beta J) / \beta^{-1} T)$, in energy density.

\end{enumerate}

\noindent All four sources of errors contribute at the same order, up to logarithmic factors.
To translate the fluctuations in the energy density to an estimate of the achievable inverse temperature, we note that $e = \mathcal{O}(1/\beta^2 J)$ at low temperatures.
This implies that the fluctations in the effective inverse temperature are given by $\delta \beta / \beta = \mathcal{O}(\delta e / e)$.
To obtain a state with effective temperature $\beta^{-1}$, we require that these relative fluctuations are small.
Focusing on the first three error sources, we find that this can be achieved with an evolution time, $T = \mathcal{O}(\beta \log(\beta J))$.
This is independent of the number of fermions $N$, and grows roughly linearly in the desired inverse temperature.

\subsection{Finite density of matter excitations}

Let us first address the generation of matter excitations.
From Section~\ref{sec: qu matter}, we know that matter excitations are produced at a rate
\begin{equation}
    \frac{ d n_m}{d t} \approx \left( \frac{ d \phi^*}{d t} \right)^2 \frac{\beta^*}{(\beta^* J)^{2 \Delta +1}},
\end{equation}
where we let $n_m$ denote the density of matter excitations (i.e.~the excitation number divided by $N$).
The numerator in the ratio corresponds to the inverse gap, and the denominator to the inverse matrix element.
Each  excitation has energy $\mathcal{O}(1/\beta^*)$, corresponding to the gap in the matter sector.

To understand the total energy of matter excitations at the end of the cooling protocol, we must understand how excitations created at early times propagate to late times.
In particular, we will assume that the Hamiltonian parameters are changed slowly enough so that the excitations propagate adiabatically: that is, a single excitation created at early times, of energy $\mathcal{O}(J)$, evolves to a single excitation at late times, of energy $\mathcal{O}(\beta^{-1})$.
Thus, the total number of excitations at the end of the protocol is simply given by the integral of the creation rate over time,
\begin{equation}
    n_m \approx \int_0^T dt \left( \frac{ d \phi^*}{d t} \right)^2 \frac{\beta^*}{(\beta^* J)^{4 \Delta +2}} \sim \frac{\log(\beta J)}{T} \int d\phi^* \frac{\beta^*}{(\beta^* J)^{4\Delta+2}} = \mathcal{O}\left( \frac{\log(\beta J)}{J T} \right),
\end{equation}
which leads to a matter excitation energy density,
\begin{equation}
    \delta e_m = \mathcal{O}( \beta^{-1} n_m ) = \mathcal{O}\left( \frac{\log(\beta J)}{(\beta J) T} \right).
\end{equation}
Taking the ratio with the energy density, $e = \mathcal{O}(1/\beta^2 J)$, we find a relative energy increase,
\begin{equation}
    \frac{\delta e_m}{e} = \mathcal{O}\left( \frac{\log(\beta J)}{\beta^{-1} T} \right).
\end{equation}
Therefore, to ensure small relative energy fluctuations, we require an evolution time of at least $T = \mathcal{O}( \beta \log(\beta J))$.

\subsection{Sudden state and stop of the ramp}

We now turn to the graviton dynamics, which are described by the semi-classical harmonic oscillator, Eq.~(\ref{eq: classical SHO}), at low temperatures.
During the cooling protocol, the center of the harmonic oscillator begins stationary at value $\phi^*_i$, then moves at constant velocity $v = \Delta \phi/T$ from $\phi^*_i$ to $\phi^*_f$, and then halts at value $\phi^*_f$.
The spring constant, $\tilde{k}(t)$ of the harmonic oscillator also changes in time as the center moves.

Since the movement in the oscillator center occurs at a constant velocity during the ramp, we are free to transform to a moving frame, $\phi' \equiv \phi - v t$, $p' = p - \tilde{m} v$, in which the center is stationary.
In this frame, the oscillator receives an instantaneous initial acceleration at time zero, from velocity zero to velocity $v$, and a instantaneous deceleration at time $T$, from velocity $v$ to velocity zero.
The initial acceleration produces a kinetic energy $\tilde{m} v^2 / 2$.
To determine how this energy propagates to later times, we leverage the notion of a \emph{classical} adiabatic invariant.
We briefly review this notion in the following paragraph, and then return to our analysis in the paragraph after.

In any a classical mechanical system with closed orbits in phase space, the classical adiabatic theorem states that the area enclosed by the orbits in phase space remains constant under a slow change in the Hamiltonian parameters~\cite{arnol2013mathematical}.
This is easily derived in the case of the harmonic oscillator.
The time-derivative of the energy  of the harmonic oscillator is,
\begin{equation}
    \dot E = \frac{1}{2} \dot{\tilde{k}} \phi'^2,
\end{equation}
where we apply Hamilton's equations of motion to simplify the expression.
Assuming that the spring constant changes slowly compared to the oscillations in $\phi'$, we can replace the $\phi'^2$ term above by its average over a single period of time-evolution, $\phi'^2 \rightarrow E/\tilde{k}$.
This yields a simple expression for the change in energy over time,
\begin{equation}
    \frac{\dot E}{E} = \frac{1}{2} \frac{\dot{\tilde{k}}}{\tilde{k}},
\end{equation}
which is easily solved to give
\begin{equation} \label{eq: classical adiabatic invariant}
    \frac{E}{\omega} = \text{constant},
\end{equation}
where $\omega = \sqrt{\tilde{k}/\tilde{m}}$ is the frequency of the oscillator.
The ratio of the energy to the frequency is thus preserved in time, and is referred to as an adiabatic invariant of the harmonic oscillator.

Returning to our analysis, we see from Eq.~(\ref{eq: classical adiabatic invariant}) that the initial energy, $\tilde{m} v^2 /2$, is damped over time by a factor $\omega_f/\omega_i$.
This results in an excess energy density, $\delta e_{a,i} \sim \tilde{m} v^2 (\omega_f/\omega_i) = \mathcal{O}(\log^2(\beta J)/\beta J^2 T^2)$, at the end of the protocol.
Meanwhile, the final deceleration produces a typical kinetic energy, $\delta e_{a,i} \sim \tilde{m} v^2 / 2 = \mathcal{O}(\log^2(\beta J)/ J T^2)$.
This is larger than the energy density associated with the initial acceleration and thus dominates the error.
Taking the ratio with the SYK energy density, $e = \mathcal{O}(1/\beta^2 J)$, we find a relative increase, $\delta e_a / e = \mathcal{O}(\log^2(\beta J)/ \beta^2 T^2)$.

\subsection{Adiabatic change in the spring constant}

We can also quantify the magnitude of corrections to classical adiabaticity, i.e.~Eq.~(\ref{eq: classical adiabatic invariant}).  
These corrections arise from the fact that the oscillator frequency, through its dependence on the spring constant, changes by a small but non-zero amount, $\delta \omega$, over the course of a single period of the harmonic oscillator.
This causes the average of $\phi'^2$ over a period to not precisely equal $E/\tilde{k}$, and thus the action to not precisely be conserved.

Surprisingly, the magnitude of these corrections is, as far as we are aware, relatively rarely discussed in literature on classical adiabatic invariants.
Nonetheless, a straightforward analysis~\cite{mohallem2018two} shows the action $J$ increases by an amount,
\begin{equation}
    \delta J \sim J \left( \frac{\delta \omega^2}{\omega^2} \right),
\end{equation}
over the course of a single period (where again, $\delta \omega$ is the change in the oscillator frequency over the period).
Transforming to continuous time, the action increases at a rate,
\begin{equation}
    \frac{1}{J} \frac{d J}{dt} \sim \omega \cdot \frac{ \left( \frac{ d \omega}{ dt } \omega^{-1} \right)^2}{\omega^2} \sim \frac{1}{\omega^3} \left( \frac{d \omega}{dt} \right)^2.
\end{equation}
In our chosen ramp, the frequency decreases exponentially in time, $d \omega / d t \sim \omega (\Delta \phi^*/ T)$.
Integrating over time, we find a total increase in the action, 
\begin{equation}
    \frac{\Delta J}{J} \sim \int_0^T dt \left( \frac{d\phi^*}{T} \right)^2 \frac{1}{\omega} = \frac{\Delta \phi}{T} \int d\phi^* \frac{1}{\omega} = \mathcal{O}\left( \frac{\log(\beta J)}{\beta^{-1} T} \right).
\end{equation}
Thus, the corrections to adiabaticity are negligible as long as the fraction in the rightmost expression is small.

\subsection{Robustness to perturbations}

We conclude by demonstrating that our estimated evolution time is robust to perturbations that might break the emergent Galilean invariance of the semi-classical graviton dynamics. 
To be concrete, we consider a simple perturbation corresponding to a frictional force\footnote{This is motivated, in part, by numerical observations that oscillations in the Maldacena-Qi model often appear to contain a small damping~\cite{lantagne2020diagnosing}. Such a damping might occur, for example, due to higher-temperature corrections to the graviton dynamics.}, $-\gamma  \omega p$.
Thus, the system evolves under the equation of motion,
\begin{equation}
    F = \dot{p} =  - \tilde{k}(t) [ \phi - \phi^*(t) ] - \gamma  \omega p,
\end{equation}
with $p \equiv \tilde{m} \dot{\phi}$.
The frictional force causes momentum in the harmonic oscillator to decay on a time-scale $\sim \! 1/\gamma \omega$.
The small parameter $\gamma$ quantifies the ratio of the oscillation time-scale, $\omega^{-1}$, to this decay time.

We perform a rough estimation of the effect of this damping on our cooling protocol as follows.
Throughout the linear ramp in $\phi^*$, there is a competition between the movement of the ramp (which favors the oscillator moving to the right with velocity $v$) and the frictional force (which favors the oscillator remaining stationary).
Roughly, we expect this competition to dissipate an amount of energy, $\tilde{m} v^2 / 2$, over a time-scale, $1/\gamma \omega$.
Since the energy of the entire coupled Hamiltonian is preserved, this dissipation must lead to an increase in energy, $\tilde{m} v^2 / 2$, among some other degrees of freedom of the Hamiltonian.
To estimate the energy at the conclusion of the cooling protocol, we need to understand how these increases in energy at intermediate times of the cooling protocol propagate to the final time $T$.
Motivated by our previous analyses of the graviton and matter excited states, we will assume that these energies propagate adiabatically: that is, an excitation energy $\tilde{m} v^2/2$ that occurs when the system has frequency $\omega$ propagates to an energy $(\omega_f / \omega) \tilde{m} v^2 /2$ when the system has frequency $\omega_f$.
Making this assumption, and taking the continuous-time limit, we find a total excitation energy,
\begin{equation}
    \delta e_p \sim \int_0^T dt \, (\gamma \omega) \left( \frac{1}{2} \tilde{m} v^2 \right) \left( \frac{\omega_f}{\omega} \right) = \gamma \omega_f \tilde{m} v^2 T = \mathcal{O}\left( \frac{\gamma \log^2(\beta J)}{\beta J T} \right).
\end{equation}
Taking a ratio with the bare energy density $e = \mathcal{O}(1/\beta^2 J)$, we find a relative increase, $\delta e_p / e = \mathcal{O}(\gamma \log^2(\beta J)/\beta^{-1} T)$.

\end{document}